\documentclass[preprint,12pt,3p]{elsarticle}
\usepackage{latexsym}
\usepackage{dcolumn}% Align table columns on decimal point
\usepackage{supertabular,multirow}
\usepackage{epsfig}
\usepackage{bm}% bold math
\usepackage{subfigure}
\usepackage{color}
\usepackage{xurl}% xurl allows for line breaks unlike hyperref
\usepackage{tikz,siunitx}
\usepackage{amssymb,amsfonts,amsmath}
\usepackage{euscript}
\usepackage{graphics}
\usepackage{setspace}
\usepackage{graphicx}
\usepackage{hyperref}% add hypertext capabilities
\usepackage{xcolor}
\usepackage{siunitx}
\usepackage[normalem]{ulem}
\usepackage[colorinlistoftodos]{todonotes}

\graphicspath{{./}, {Figs/}}

\definecolor{darkgreen}{rgb}{0, 0.5, 0.05}

\newcommand{\deleted}[1]{}

\newcommand{\eqn}[1]{\begin{align}#1\end{align}}
\newcommand{\bs}[1]{\boldsymbol{#1}}

\newcommand{\pare}[1]{\left( #1 \right) }
\newcommand{\corchete}[1]{\left[ #1 \right]}
\newcommand{\fr}[2]{\frac{#1}{#2}}
\newcommand{\wtil}[1]{\widetilde{#1}}
\newcommand{\mc}[1]{\mathcal{#1}}
\newcommand{\avg}[1]{\langle #1 \rangle}
\newcommand{\tex}[1]{\mbox{\scriptsize{#1}}}

\def\dd{\mathrm{d}}  
\def\kt{k_B T}
\def\bna{\bs{\nabla}}

\def\bA{\bs{A}}
\def\bc{\bs{c}}

\def\bD{\bs{D}}
\def\be{\bs{e}}
\def\bbf{\bs{f}}
\def\bF{\bs{F}}

\def\bG{\bs{G}}
\def\bI{\bs{I}}

\def\bL{\bs{L}}

\def\bZ{\bs{Z}}
\def\bM{\bs{M}}
\def\bN{\bs{N}}
\def\bn{\bs{n}}

\def\bP{\bs{P}}
\def\bK{\bs{K}}
\def\bq{\bs{q}}

\def\br{\bs{r}}

\def\bS{\bs{S}}

\def\bT{\bs{T}}
\def\bu{\bs{u}}
\def\bU{\bs{U}}
\def\bv{\bs{v}}
\def\bV{\bs{V}}
\def\bw{\bs{w}}
\def\bW{\bs{W}}
\def\bx{\bs{x}}

\def\by{\bs{y}}

\def\bzero{\bs{0}}

\def\btheta{\bs{\theta}}

\def\bSigma{\bs{\Sigma}}
\def\btau{\bs{\tau}}
\def\bomega{\bs{\omega}}

\def\blambda{\bs{\lambda}}
\def\balpha{\bs{\alpha}}
\def\bk{\bs{k}}

\def\bmS{\bs{\mc{S}}}
\def\bmD{\bs{\mc{D}}}
\def\bmL{\bs{\mc{L}}}
\def\bmT{\bs{\mc{T}}}

\def\mcB{\mc{B}}
\def\bmG{\bs{\mc{G}}}

\def\bmZ{\bs{\mc{Z}}}
\def\bmI{\bs{\mc{I}}}

\begin{document}

\title{A scalable method to model large suspensions of colloidal phoretic particles with arbitrary shapes}

\author[mymainaddress]{Blaise Delmotte}
\author[mysecondaryaddress]{Florencio Balboa Usabiaga}

\address[mymainaddress]{LadHyX, CNRS, Ecole Polytechnique, Institut Polytechnique de Paris, 91120 Palaiseau, France}
\address[mysecondaryaddress]{BCAM - Basque Center for Applied Mathematics, Mazarredo 14, Bilbao, E48009, Basque Country - Spain}

\date{\today}

\begin{abstract}
  % Colloidal phoretic particles with complex shapes are increasingly used to perform specific functions, such as pumping, mixing or guided transport, in Fluid Mechanics, Soft Matter and Engineering in general.
  Phoretic colloids self-propel thanks to surface flows generated in response to surface gradients (thermal, electrical, or chemical),
  that are self-induced and/or generated by other particles.  
  % The dynamics of phoretic suspensions thus results from the interplay between phoretic and hydrodynamic interactions between the suspended particles.
  % In this work we focus on diffusio-phoretic particles that move in response to chemical concentration gradients.
  % While many efficient numerical methods already exist for solving the hydrochemical interactions between spherical phoretic particles,
  % the current techniques at hand for more complex shapes are expensive and limited to one or less than ten particles in dynamic simulations. 
  Here we present a scalable and versatile framework to model chemical and hydrodynamic interactions in large suspensions of arbitrarily shaped phoretic particles,
  accounting for thermal fluctuations at all Damkholer numbers.
  Our approach, inspired by the Boundary Element Method (BEM), employs second-layer formulations,
  regularised kernels and a grid optimisation strategy to solve the coupled Laplace-Stokes equations
  with reasonable accuracy at a fraction of the computational cost associated with BEM.
  As demonstrated by our large-scale simulations, the capabilities of our method enable the exploration of new physical phenomena that,
  to our knowledge, have not been previously addressed by numerical simulations.
\end{abstract}
\begin{keyword}
Reactive particles \sep  Phoresis \sep Stokes flow \sep Suspensions \sep Colloidal particles \sep Active matter \sep Complex fluids \sep Large scale simulations
% %% PACS codes here, in the form: \PACS code \sep code
% %% MSC codes here, in the form: \MSC code \sep code
% %% or \MSC[2008] code \sep code (2000 is the default)
\end{keyword}
\maketitle
\tableofcontents

\section{Introduction}
\label{sec:Intro}
   The individual and collective dynamics of self-propelled phoretic particles have attracted significant attention in recent decades  \cite{MoranPosner2017,Stark2018,Illien2017,Dominguez2022,Zottl2023}. These particles move with phoretic flows generated on their surface in response to surface gradients (thermal, electrical, or chemical), which are self-induced and/or generated by other particles \cite{Marbach2019}. The most common  realizations of such systems involve spherical and rod-like particles, for which a plethora of individual and collective phenomena have been documented both theoretically and experimentally. Examples include rheotaxis \cite{ Palacci2015,Ren2017,Katuri2018,Brosseau2019,Sharan2022}, gravitaxis \cite{Campbell2013,Ten2014,Brosseau2021}, dynamical clustering and self-assembly \cite{TheurkauffAllBocquet2012,GinotAllCottinBizonne2018, Palacci2013, Varma2018, Singh2019,Pohl2014,Wykes2016,Liebchen2017}. Recent studies have explored particles with more intricate shapes, such as fore-aft asymmetric catalytic particles \cite{Shklyaev2014, Michelin2015} used for the collection and degradation of microplastics \cite{Chattopadhyay2023}, rotating chiral particle \cite{Brooks2019,Sharan2021} and stirrers \cite{Zhang2019,Kumar2022},  phoretic wind turbines \cite{Shen2019} and pumps \cite{Michelin2015b,Tan2019}, shaped-programmed microtori for particle transport \cite{Baker2019}, L-shaped swimmers following circular orbits \cite{Kummel2013},  tadpole-shaped catalytic swimmers with shape-programmed trajectories \cite{lv2020,Mu2022},  phoretic fibers and sheets whose motion is modulated by dynamic deformations \cite{Katsamba2020,Montenegro2018,Laskar2019,Manna2022}, and diamond-like photocatalytic particles \cite{Heckel2022}.\\
   
   The emerging motion and patterns reported in these works result from the interplay between phoretic and hydrodynamic interactions between the suspended particles and the bounding walls. In the following we will mainly focus on self-diffusiophoretic motion, i.e. due to concentration gradients of solute particles that result from surface chemical reactions. Nonetheless, the framework presented below also applies to thermophoresis (temperature gradients), dielectrophoresis or  induced-charge electrophoresis (electric potentials). 
   Owing to the small particle size ($O(1 \mu$m$)$) and the magnitude of the phoretic flows on their surface, the Péclet number characterizing the motion of the solute particles is small enough to neglect advection and consider only the diffusive part. The corresponding Reynolds number is also small enough to neglect inertial and unsteady terms in the Navier-Stokes equations governing the flow around the particles. As a result, modelling the hydrochemical interactions at play in these systems requires solving sequentially two problems: a Laplace problem for the solute concentration field, subject to boundary conditions dictated by the chemical reactions on the particle and wall surfaces, and then a Stokes problem for the particle velocities, subject to phoretic slip surface velocities induced by the solute gradients obtained from the Laplace problem. \\
   
    Both problems, being elliptic, have an integral representation from which a multipolar expansion can be derived.  
    Low order multipolar expansions are typically used in far-field models, where the particle is represented by a series of singularities distributed at its center. The magnitude of the singularities are obtained from the boundary conditions on the particle surface. In the simplest models, only the slowest decaying singularities are accounted for and their strength is assumed to be constant, as if the particle was isolated \cite{Pohl2014,LiebchenLowen2019,KansoMichelin2019}. Mutual interactions can be accounted for when computing the magnitude of these singularities, which is the case of the method of reflections \cite{VarmaMichelin2019}. These methods are mostly reliable in the far-field and do not account for the particle finite size and geometry. 
    For spherical objects, the finite size of particles can be accounted for in truncated multipolar expansions by using Faxen's laws, to obtain the singularity strengths, and including  higher order moments which involve the particle radius, as done in Stokesian Dynamics \cite{BradyBossis1988}, Laplacian Dynamics \cite{Yan2016a} and tensorial spherical harmonics \cite{Singh2019,Singh2019pystokes}. Alternatively, the finite particle size can be accounted for with regularized  kernels whose size is related to the particle radius, as in the Method of Regularized Stokeslets \cite{Cortez2005}, the Force Coupling Method \cite{MaxeyPatel2001,Rojas-Perez2021} or Immersed Boundary Methods \cite{Bhalla2013}. The Method of Regularized Stokeslets is based on the Green's functions of the Stokes problem, while Immersed Boundary Methods interpolate the flow computed on a grid. The Force Coupling Method works with both approaches \cite{YeoMaxey2010,Rojas-Perez2021,Su2023}. \\
    
    At the other side of the spectrum, the  Boundary Element Method (BEM) uses a direct discretization of the integral formulation of the Laplace and Stokes problems \cite{Pozrikidis1992,Pozrikidis2002}. Thanks to its high accuracy, this method has been widely used to study the motion of isolated diffusio-phoretic particles with various shapes \cite{Poehnl2021,Simmchen2016,Uspal2015}.  However, despite recent efforts to improve its scalability, BEM remains costly and  cannot include thermal fluctuations in 3D. It also faces several numerical challenges due to the  singular nature of the boundary integral operators (BIO). In the case of spherical particles, \cite{Corona2018,Kohl2023} diagonalized the BIO on the sphere surface with spherical harmonics, which allows to circumvent the discretization of singular integrals and efficiently evaluate self and near-field interactions \cite{Yan2020b}. In order to avoid handling singularities in the self and near-field interactions \cite{Montenegro-Johnson2015} used the regularized Stokeslet kernel in the BIO. This allowed them to simulate phoretic particles with various shapes in conjunction with experiments \cite{Baker2019} and theory \cite{Katsamba2020, Varma2018}.\\
    
    While quite a few methods exist to solve the hydrochemical interactions between spherical particles in an efficient and scalable way, the current techniques at hand for non-spherical objects are expensive and limited to one or less than ten particles in dynamic simulations.  Moreover none of these methods currently include thermal fluctuations, which might be important in these colloidal systems.
     It also worth noting that many of the cited studies focused on simplified models where the surface fluxes are assumed to be independent of the solute concentration (which might not be relevant at finite or zero Damkholer numbers).\\
     
     In this work we propose a simple and flexible framework to model hydrochemical and steric interactions in large suspensions of phoretic particles with arbitrary shapes accounting for  thermal fluctuations at all Damkholer numbers. Our method is directly inspired from BEM for the Laplace problem, and utilizes the rigidmultiblob (RMB) framework \cite{Usabiaga2016} for the hydrodynamic part. In order to optimize the performances of the method, we develop a grid optimization algorithm,  inspired from \cite{Broms2023}, which matches the exact hydrodynamic response of slip-driven particles with coarse grids. These optimized coarse grids provide a 2-3 digits accuracy at a very small cost compared to BEM, which permits the simulation of large collections of  phoretic particles of arbitrary shapes. Our approach therefore offers a novel alternative bridging the gap between far-field models and BEM. Far-field models are cost-effective but lack accuracy at moderate and short distances, whereas BEM is accurate across all distances but incurs high costs, even in the far-field. Our method achieves accuracy  at moderate and short  distances, all at a fraction of the cost associated with BEM. \\
     
     We first introduce  the equations that govern the solute concentration, the flow field and the motion of the suspended particles in Section \ref{sec:Equations}. Then we write their integral forms and detail our method to discretize them with the RMB method, including thermal fluctuations. In addition, we improve the current RMB to better account for slip flows on the particle surfaces using a double layer formulation \cite{Smith2021}. In Section \ref{sec:Grid_opt} we propose a grid optimization procedure to match the exact hydrodynamic response of a single rigid particle of arbitrary shape with surface slip. We validate our method in Section \ref{sec:Validations} against analytical and numerical reference solutions from the literature.  We find that even with a low resolution, as low as 42 nodes on the particle surface, our method with optimized grids matches reference solutions for isolated particles, and achieves close agreement in the near field. 

     % The method converges to the reference solution at higher resolutions.
     Using the capacities offered by our new framework,  we turn to the simulation of larger and more complex systems in Section \ref{sec:Simulations}. First we study the collective swimming of  phoretic and Brownian Janus rods  on an incline. It has been shown that a single particle can exhibit gravitaxis by tilting against the incline and swimming uphill against very steep slopes \cite{Brosseau2021}. Here we investigate the effect of hydrochemical interactions on a whole suspension. We find that hydrochemical attraction suppresses gravitaxis and observe the emergence of small clusters that rearrange dynamically due to thermal fluctuations. These clusters rotate, translate or remain still, while sedimenting along the incline.  The second simulation consists of a large collection of chiral particles with uniform surface properties confined by a harmonic potential above a no-slip wall. Thanks to its chirality, a single particle rotates by itself due to self-phoresis \cite{Brooks2019}. We show that a large collection of such particles exhibit nontrivial collective behaviours, such as the coexistence of an outer rim and an inner crystal rotating in opposite directions.
     Our framework can readily incorporate other types of phoresis, such as thermophoresis and electrophoresis.  It  has been implemented in a collaborative code on GitHub (\url{https://github.com/stochasticHydroTools/RigidMultiblobsWall}) which is user-friendly and publicly available.

\section{Governing equations and model}
\label{sec:Equations}

\subsection{Governing equations}
\label{sec:eqns}

In this work we consider a fluid domain with a dissolved solute with concentration $c=c(\br)$ and, additionally,
$M$ rigid colloids, $\{\mcB_m\}_{m=1}^M$, immersed in the fluid.
We assume that the dynamics of the solute is dominated by diffusion, i.e.\ zero Péclet number.
Thus, in the fluid domain the concentration diffuses according to the Laplace equation 
\eqn{
  \label{eq:Laplace}
  D \bna^2 c = 0,
}
with diffusion coefficient $D$.

The immersed colloids activate chemical reactions on their surfaces to produce or consume solute.
Such reactions are introduced as boundary conditions of the Laplace equation.
The boundary conditions are imposed on the concentration fluxes, i.e.\ Robin boundary conditions,
and to simplify the mathematical problem we only consider linear boundary conditions.
In particular, we consider sinks that consume solute at a rate proportional to the local concentration,
sources that produce solute at a constant flux or a linear combination of both \cite{Lu1998, MichelinLauga2014,Rojas-Perez2021}
\eqn{
  \label{eq:BC}
  D \bn \cdot \bna c = kc - \alpha \; \text{ on } \partial \mcB_m,
}
where $\bn=\bn(\br)$ is the surface normal on the colloids, $k=k(\br)$ the surface reaction rate and $\alpha=\alpha(\br)$ the surface production flux,
both of which can vary over the colloidal surface.
The minus sign on front of the production flux is included for convenience so positive values of $\alpha$ indicate production of solute.
Since the reaction flux is proportional to the local concentration while the concentration diffuses with a finite diffusion coefficient $D$, this
boundary condition allows to model either diffusive-limited or reaction-limited systems.
The balance between diffusion and reaction is characterized by the
the Damköhler number, $\text{Da} = k R / D$, which represents the ratio between the diffusive time over one colloidal radius, $R^2/D$, and the reaction time scale $R/k$.
Thus, the Damköhler number is large for diffusive-limited systems and small for reaction limited systems \cite{Bhalla2013}.
We verify in Sec.\ \ref{sec:damkholer} that our approach works for arbitrary Damköhler numbers.
Additional boundary conditions at infinity must be included.
In general, we will consider two scenarios: an unbounded domain and a half-space above an infinite plane wall.
In the first case we will use the boundary condition $c \xrightarrow[{r \rightarrow \infty}]{} c_{\infty}$.
In the second case we will also assume a zero flux of concentration through the wall.

The fluid domain not only diffuses the solute but also carries the hydrodynamic interactions between colloids.
The fluid flows are modeled as viscous dominated, and thus governed by the Stokes equations \cite{Pozrikidis2002}
\eqn{
\label{eq:Stokes}
- \nabla p + \eta \nabla^2 {\bs{v}} &= \bna \cdot \bmZ, \\
\nabla \cdot {\bs{v}} &= 0,
}
where $\bv$ and $p$ are the flow velocity and pressure and $\eta$ the fluid viscosity.
The right hand side in \eqref{eq:Stokes} includes the divergence of a stochastic stress tensor $\bmZ$ responsible for the thermal fluctuations or Brownian motion \cite{Landau1987,Fabritiis2007}.
The stochastic stress tensor is delta correlated in space and time
\eqn{
  \avg{\mc{Z}_{ij}(\bx, t) \mc{Z}_{kl}(\bx', t^{'})} = 2\eta \kt \pare{\delta_{ik}\delta_{jl} + \delta_{il}\delta_{jk}} \delta(\bx- \bx') \delta(t - t'),
}
and the correlations magnitude depends on the thermal energy $\kt$.

On the colloidal surfaces the fluid obeys a \emph{slip} condition \cite{MichelinLaugaBartolo2013, Rojas-Perez2021}
\eqn{
  \label{eq:slip-cond}
  \bv(\br) = \bu_m + \bomega_m \times (\br - \bq_m) + \bu_s(\br)  \; \text{ on } \partial \mcB_m.
}
The first two terms in the right hand side of \eqref{eq:slip-cond} represent the rigid motion of colloid $m$
with linear and angular velocities $\bu_m$ and $\bomega_m$ with respect to its tracking point (e.g.\ its center) $\bq_m$.
The last term in \eqref{eq:slip-cond} represent the slip introduced by the phoretic effects of the concentration near the colloid surface.
The slip velocity is proportional to the surface concentration gradient \cite{Anderson1989,Marbach2019}
\eqn{
  \label{eq:slip}
  \bu_s = \mu \bna_{\parallel} c = \mu \pare{\bI - \bn\bn^T} \bna c \; \text{ on } \partial \mcB_m,
}
where $\mu=\mu(\br)$ is the surface mobility and $\bna_{\parallel}$ the surface gradient,
i.e.\ the gradient projected to the surface tangent plane.

The equations are closed with the balance of force and torque: 
the hydrodynamic stress on the colloids equates the rest of external forces and torques, $\bbf_m$ and $\btau_m$, acting on them \cite{Pozrikidis1992}
\eqn{
  \label{eq:balanceF_continuum}
  \int_{\partial \mcB_m} \blambda(\br) \,\dd S_{\br} = \bbf_m, \\
  \label{eq:balanceT_continuum}
  \int_{\partial \mcB_m} (\br - \bq_m) \times \blambda(\br) \,\dd S_{\br} = \btau_m,
}
where $\blambda$ is a single layer potential that enforce the slip condition and the balance of force and torque.
The equations \eqref{eq:Laplace} and \eqref{eq:BC} form a linear system for the concentration.
Once the surface slip is known through \eqref{eq:slip} the equations \eqref{eq:Stokes},\eqref{eq:slip-cond},\eqref{eq:balanceF_continuum} and \eqref{eq:balanceT_continuum}
form a linear system for the single layer potential $\blambda$ and the colloidal velocities $\{\bu_m, \bomega_m\}_{m=1}^{M}$.

\subsection{Integral formulation of the Laplace equation}
\label{sec:integral_Laplace}

Since both the Stokes and the Laplace equation are elliptic partial differential equations it is possible to solve them without discretizing the whole fluid domain but only
its boundaries, i.e.\ the colloid surfaces, by using an integral formulation \cite{Montenegro-Johnson2015}.
Before we present the integral formulation we introduce the Laplace Green's function and its derivatives in an unbounded domain.
The Green's function of the Laplace equation is
\eqn{
  \label{eq:kernel_monopole}
  G(\bx, \by) = \fr{1}{4\pi} \fr{1}{r},
}
where $\by$ and $\bx$ are the source and target (observation point) respectively and $r=|\br| = |\bx - \by|$.
Taking the derivative with respect to the source, $\by$, we get the dipole kernel
\eqn{
  \label{eq:kernel_dipole}
  T_i(\bx, \by) = \fr{\partial}{\partial y_i} G(\bx, \by) &= \fr{1}{4\pi}\fr{r_i}{r^3}.
}
Taking the derivative of $T_i(\bx,\by)$ with respect to the target, $\bx$, we get the quadrupole
\eqn{
  \label{eq:kernel_quadrupole}
  L_{ij}(\bx,\by) = \fr{\partial}{\partial x_j} T_i(\bx, \by) &= \fr{1}{4\pi} \fr{\delta_{ij}}{r^3} - \fr{3}{4\pi}\fr{r_i r_j}{r^5}.
}

Using these kernels we introduce the single layer and double layer operators
\eqn{
  \label{eq:single_layer_op}
  \bmS[c](\bx) &= \int_{\partial \mcB} G(\bx-\by) c(\by) \,\dd S_y, \\
  \label{eq:double_layer_op}
  \bmD[c](\bx) &= \int_{\partial \mcB} T_i(\bx-\by)n_i(\by) c(\by) \,\dd S_y, 
}
where the integral is over all the colloidal surfaces: $\partial \mcB = \bigcup_{m=1}^{M}\partial \mcB_m$.
We also introduce two auxiliary operators 
\eqn{
  \label{eq:aux1_op}
  \bmL_i[c](\bx) &= \int_{\partial \mcB} L_{ij}(\bx-\by) n_j(\by) c(\by)  \,\dd S_y, \\
  \label{eq:aux2_op}  
  \bmT_i[c](\bx) &= \int_{\partial \mcB} T_i(\bx-\by)  c(\by) \,\dd S_y. 
}

The integral formulation for the Laplace equation  can be written as \cite{Montenegro-JohnsonMichelinLauga2015}
\eqn{
  \label{eq:complete_v1}
  \corchete{\fr{1}{2}\bmI - \bmD} c = -\bmS\corchete{\fr{\partial c}{\partial \bn}} + c_{\infty} \;\; \text{ on } \partial \mcB,
}
where the double layer operator $\bmD$ acts in the principal value sense and $\bmI$ is the identity operator \cite{Pozrikidis1992}.
The  normal derivative of the concentration, $\partial c / \partial \bn$, on the right hand side can be replaced with the boundary conditions on the colloids, Eq.\ \eqref{eq:BC}, to rewrite the equation as 
\eqn{
  \label{eq:complete_v2}
  \corchete{\fr{1}{2}\bmI - \bmD + \bmS\corchete{\fr{k}{D} }} c = \bmS\corchete{\fr{\alpha}{D}}  + c_{\infty}.
}
Since the surface parameters $k=k(\br)$ and $\alpha=\alpha(\br)$ define the boundary conditions,
we have absorbing boundary conditions for $(k>0,\,\alpha=0)$ and emitting boundary conditions for $(k=0,\,\alpha>0)$.
In general, it is possible to use $k,\alpha>0$ to model a colloidal suspension of mixed sources and sinks.

Once the concentration is known, its gradient at the surfaces can be computed with the derivative of \eqref{eq:complete_v2}, i.e.\
\eqn{
  \label{eq:grad_c}
  \fr{1}{2}\bna c =  \bna c_{\infty} + \bmL[c] - \bmT\corchete{-\fr{k}{D}c + \fr{\alpha}{D} },
}
and then, we can use \eqref{eq:slip} to compute the active slip and solve the Stokes equations.
Additionally, we can compute the concentration anywhere in the fluid domain with
\eqn{
  c(\bx) = c_{\infty}  +  \bmD[c] - \bmS \corchete{\fr{k}{D}c - \fr{\alpha}{D}}.
}

\subsection{Integral formulation of the Stokes equations}
\label{sec:integral_Stokes}

The Stokes equations also admit an integral formulation.
We start with the second layer formulation \cite{Pozrikidis1992, Smith2021}  
\eqn{
  \label{eq:double_layer}
  \pare{\fr{1}{2}\bmI + \bmD_{\tex{St}}} \bv(\br) + \bu_{\text{th}} = (\bmG_{\text{St}} \blambda)(\br) \;\;\; \text{for }\br \text{ on } \partial \mcB_m,
}
where $\bmG_{\text{St}}$ is the single layer operator of the Stokes equation acting on the single layer potential $\blambda$
and $\bmD_{\tex{St}}$ is the Stokes double layer operator \cite{Pozrikidis1992}.
In an unbounded domain 
\eqn{
  \label{eq:operators_Stokes}
  \pare{\bmG_{\text{St}} \blambda}_i(\bx) = \int_{\partial \mcB} \fr{1}{8\pi \eta r} \pare{\delta_{ij} + \fr{r_i r_j}{r^2}} \lambda_j(\by) \; \dd S_y, \\
  \pare{\bmD_{\tex{St}} \bv}_i(\bx) = \int_{\partial \mcB} -\fr{3}{4\pi} \fr{r_i r_j r_k}{r^5} n_k(\by) v_j(\by) \; \dd S_y,
}
where $\br = \bx - \by$ and $\bn=\bn(\by)$ is the surface normal.
The new term in \eqref{eq:double_layer}, $\bu_{\text{th}}(\br)$, represents the thermal fluctuations introduced by the stochastic stress.
Replacing the slip condition on the particle surface, $\bv(\br) = \bu_m + \bomega_m \times (\br - \bq_m) + \bu_s(\br)$,
in \eqref{eq:double_layer} we get the formulation
\eqn{
  \label{eq:slip_integral}
  \bu_m + \bomega_m \times (\br - \bq_m) + \pare{\fr{1}{2}\bmI + \bmD_{\tex{St}} }\bu_s(\br) + \bu_{\text{th}}(\br) = (\bmG_{\text{St}} \blambda)(\br) \;\;\; \text{for }\br \text{ on } \partial \mcB_m,
}
where we have used the fact that the double layer operator acting on rigid body velocities is equivalent to half the identity operator \cite{Pozrikidis1992, Smith2021}.

This equation can be simplified further.
 Since the flow rate of the prescribed boundary velocity \eqref{eq:slip-cond} across the particle surfaces is zero, i.e.\ $\int_{\partial \mcB} \bv(\br)\cdot\bn(\br)\dd S = 0$,
  then one can extend the incompressible flow continuously inside the bodies and eliminate the double layer potential \cite{Pozrikidis1992,Smith2021}. The double layer (DL) formulation  can therefore be simplified to a single layer (SL) formulation  where  the expression  $\pare{\fr{1}{2}\bmI + \bmD_{\tex{St}} }\bu_s(\br)$ in \eqref{eq:slip_integral} is replaced by  $\bu_s$. With the SL formulation, the single layer potential $\blambda(\br)$ does not correspond to the hydrodynamic tractions, but to a potential that enforces the slip condition and the force and torque balance \cite{Usabiaga2016, Smith2021}. 
The difference between the DL and SL formulation therefore lies in the way  the slip term is treated, $\pare{\fr{1}{2}\bmI + \bmD_{\tex{St}} }\bu_s(\br)$ for DL \textit{vs.} $\bu_s(\br)$ for SL, and in the interpretation of the single layer potential $\blambda$. 
While these two formulations are \textit{strictly equivalent} in the continuous setting, Smith \textit{et al.} \cite{Smith2021} have found that, in the discrete setting using Regularized Stokeslets, the DL formulation is 3 to 4 times more accurate than SL. As shown in Sec.\ \ref{sec:valid_slip}, we also find that the error is reduced with DL.

% The zero contribution of the phoretic slip to the total force on the colloids permits to eliminate the double layer operator
% from the equation using $\pare{\fr{1}{2}\bmI + \bmD_{\tex{St}} }\bu_s(\br) = \bu_s(\br)$ \cite{Smith2021}. 
% However, we found that, in the discrete setting, the numerical error is lower if we keep the second-layer operator, $\bmD_{\tex{St}}$, as we show in Sec.\ \ref{sec:valid_slip} (see also \cite{Smith2021}).
The stochastic slip is equivalent to including a stochastic stress tensor directly in the Stokes equations as in \eqref{eq:Stokes}
provided it has the appropriate covariance \cite{Delong2014a, Sprinkle2019}.
In Sec.\ \ref{sec:disc_Stokes} we explain how to generate the fluctuations with the right covariance.
The slip condition is supplemented with the balance of force and torque, Eqs.\ \eqref{eq:balanceF_continuum}-\eqref{eq:balanceT_continuum}, to close the linear system.
% We note in passing that, in absence of slip terms, the single layer potential, $\blambda$, is the hydrodynamic traction, but in general it is just a potential that enforce the slip condition and the force and torque balance \cite{Usabiaga2016, Smith2021}.

One can use the same formalism to solve the hydrochemical problem above an infinite impermeable wall by using the Laplace and Stokes kernels in half-space.
In \ref{sec:green} we provide the kernels of the Laplace equation in half space, i.e.\ above an infinite no-flux wall while
the Stokes kernels can be found, for example, in Refs.\ \cite{Swan2007,Mitchell2017}.\\

\subsection{Discretization of the Laplace problem}
\label{sec:disc_Laplace}

In these sections we discuss the discretization to solve the Laplace and Stokes problems.
We use the same grid to solve both the Stokes and the Laplace equations.
This way, the gradients on the particle surface can be used directly to compute the active slip appearing on the Stokes equations.
First, we discretize the surface of the rigid bodies with nodes (also called blobs) located at positions $\br_i$, $i=1,\cdots,N_b$, where $N_b$ is  the total number of nodes in the system.
The optimal position for the nodes is discussed in Sec.\ \ref{sec:Grid_opt}.
Then, using the nodes as collocation points to solve the discretization of \eqref{eq:complete_v2}, we obtain the linear system 
\eqn{
  \label{eq:complete_discrete}
  % \sum_{j=1}^{N_b} \corchete{\fr{1}{2 w_j}\bI_{ij} - \bD_{ij} + \bS_{ij}\corchete{\fr{k_j}{D} \cdot }} c_j w_j = \sum_{j=1}^{N_b} \bS_{ij}\corchete{\fr{\alpha_j}{D}} w_j  + c_{\infty}(\br_i) \;\; \mbox{for all } i, \\
  \corchete{\fr{1}{2} \bI \bw^{-1} + \bD + \bS \fr{\bk}{D}} \bw \bc = \bS \fr{\bw \balpha}{D} + \bc_{\infty},
}
where $\bI$, $\bS$ and $\bD$ are $N_b \times N_b$ matrices corresponding to the discrete versions of the continuous operators introduced in \eqref{eq:complete_v2}. $\bc$ and $\balpha$ are $N_b \times 1$ vectors collecting the value of the concentration and the reaction flux at each node.
% In this expression we arrange the value of the concentration and the reaction flux at each node in the $N_b \times 1$ vectors $\bc$ and $\balpha$ respectively.
Meanwhile, the reaction rate and the quadrature weights are arranged in the $N_b \times N_b$ diagonal matrices $\bk$ and $\bw$.
The weight values $\bw$  represent the surface area covered by the nodes.
Similarly, once the concentration vector $\bc$ is obtained, we compute the concentration gradient on the nodes with
\eqn{
  \label{eq:grad_discrete}
  % \fr{1}{2}(\bna c)_i =  (\bna c_{\infty})_i + \sum_{j=1}^{N_b} \pare{\bL_{ij}[c_j] - \bT_{ij} \corchete{-\fr{k_j}{D}c_j + \fr{\alpha_j}{D} }} w_j \;\; \mbox{for all } i. \\
  \fr{1}{2} \bna \bc = \bna \bc_{\infty} + \bL \bw \bc + \bT \corchete{\fr{\bw \bk \bc}{D} - \fr{\balpha}{D}},
}
where $\bL$ and $\bT$ are $3N_b \times N_b$ matrices corresponding to the discrete versions of the continuous operators introduced in \eqref{eq:grad_c}.
The Laplace kernels diverge when the source and target point coincide.
We do not introduce any kernel regularization to deal with this divergence, instead,
we set the kernels diagonal terms to zero in both \eqref{eq:complete_discrete} and \eqref{eq:grad_discrete}.
We will study the accuracy of this discretization in Section  \ref{sec:Validations}.
To solve the linear system \eqref{eq:complete_discrete} we use an unpreconditioned GMRES solver,
and we show that the GMRES convergence is robust and scales well with the number of colloids in Sec.\ \ref{sec:convergence}.

\subsection{Discretization of the Stokes problem }
\label{sec:disc_Stokes}

To solve the Stokes equations we use the rigid multiblob method with the same grid used to solve the Laplace equation \cite{Usabiaga2016}.
The rigid multiblob method uses simple quadrature rules to discretize the integrals over the bodies' surfaces.
Thus, the balance of force and torque become sums over the forces acting on each node
\eqn{
  \label{eq:balanceF}
  \sum_{i \in \mcB_m} \blambda_i = \bbf_m, \\
  \label{eq:balanceT}  
  \sum_{i \in \mcB_m} (\br_i - \bq_m) \times \blambda_i = \btau_m.
}
Note that here the quadrature weights are included in the values of $\blambda_i$ which now represent finite forces and not density forces as in the continuum formulation.
The slip condition, as in collocation methods \cite{Pozrikidis1992}, is evaluated at each node 
\eqn{
  \label{eq:RBM_slip_discrete}
  \sum_{j=1}^{N_b} \bM_{ij} \blambda_j &= \bu_m + \bomega_m \times (\br_i - \bq_m) + \sum_{j=1}^{N_b} \pare{\fr{1}{2}\bI_{ij} + \bD_{\tex{St},ij}} \bu_{s,j} + \bu_{\text{th},i} \;\; \mbox{for } i \in \mcB_m, 
}
where the discretized, $3N_b\times 3N_b$, mobility matrix $\bM$ mediates the hydrodynamic interactions between nodes.

The rigid multiblob method uses a regularized version of the Stokes Green's function.
The regularization is done by a double convolution of the Stokes Green's function, $\bG_{\text{St}}(\br, \br')$, with Dirac delta functions defined on the surface of a sphere of radius $a$
\eqn{
  \label{eq:RPY}
  \bM_{ij} = \bM(\br_i, \br_j)  &=  \fr{1}{(4\pi a^2)^2}  \int \delta(|\br' - \br_i|-a) \bG_{\text{St}}(\br', \br'') \delta(|\br'' - \br_j|-a) \dd^3 r'' \dd^3 r'.
}
This regularization is known as the Rotne-Prager or RPY approximation \cite{Rotne1969, Wajnryb2013}.
The advantage of the RPY mobility matrix is that it never diverges and is always positive definite, even when spheres overlap.
Thus, the numerical method is quite robust and easy to implement as it is not necessary to consider sophisticated quadrature rules to deal with the divergence of the Stokes kernel.
Moreover, the positive definiteness of the mobility simplifies the generation of the stochastic noise as we show shortly.
The same regularization is applied to the second-layer operator $\bD_{\tex{St}}$.

The active slip is evaluated independently on each node
\eqn{
  \label{eq:slip_discrete}
  \bu_{s,i} = \mu_i \pare{\bna_{\parallel} c}_i = \mu_i \pare{\bI - \bn_i\bn_i^T} \pare{\bna c}_i \; \text{ for all } i \in \mcB_m,
}
where $\mu_i$ and $\bn_i$ are the phoretic mobility and the surface normal defined at the node $i$.

The stochastic slip  is generated with the right covariance to correctly reproduce the Brownian motion.
We prove hereafter that one can use the stochastic slip 
\eqn{
  \label{eq:u_th} 
  \bu_{\text{th}} = \sqrt{2 \kt} \bM^{1/2} \bZ,
}
where $\bM^{1/2}$ is the \emph{square root} of the mobility matrix, i.e.\ $\bM^{1/2}\pare{\bM^{1/2}}^T=\bM$, and $\bZ$ is a white noise vector defined on the nodes
with covariance $\avg{\bZ(t) \bZ^T(t')} = \bI \delta(t - t')$.
This form of the noise term is equivalent to include a stochastic stress directly into the Stokes equation as in  \eqref{eq:Stokes} \cite{Delong2014a, Sprinkle2019}.
Since the mobility is positive definite, the stochastic velocity \eqref{eq:u_th} is well defined. 
Additionally, the action of the square root of the mobility can be computed with the Lanczos scheme,
a matrix free iterative method  that only requires the action of the mobility \cite{Ando2012}.

Once the active and the stochastic slips are computed, the linear equations \eqref{eq:balanceF}-\eqref{eq:RBM_slip_discrete} can be solved to find the colloidal velocities.
To display the structure of the linear system it is convenient to introduce the $3N_b \times 6M$ block diagonal matrix $\bK$ (defined implicitly in \eqref{eq:RBM_slip_discrete}) with blocks
\eqn{
  \bK_{m,i} =
  \begin{cases}
    \corchete{\bI_{3\times 3} \;\; -(\br_i - \bq_m)^{\times}} & \text{if } i \in \mcB_m, \\
    \corchete{\bzero_{3\times 3} \;\;  \bzero_{3\times 3}} & \text{otherwise},
  \end{cases}
}
where $\bI_{3\times 3}$ is the $3\times 3$ identity matrix and $(\br_i - \bq_m)^{\times} \bx = (\br_i - \bq_m) {\times} \bx$ for any vector $\bx$.
Then, the equations \eqref{eq:balanceF}-\eqref{eq:RBM_slip_discrete} can be rearranged as
\eqn{
  \label{eq:linear}
  \begin{bmatrix}
    \bM & -\bK \\
    -\bK^T & \bzero
  \end{bmatrix}
  \begin{bmatrix}
    \blambda \\
    \bU
  \end{bmatrix}
  =
  \begin{bmatrix}
    \pare{\fr{1}{2}\bI + \bD_{\tex{St}}} \bu_s + \bu_{\text{th}}\\
    -\bF
  \end{bmatrix},
}
where the vectors $\bU=\{\bu_m, \bomega_m\}_{m=1}^{M}$ and $\bF=\{\bbf_m, \btau_m\}_{m=1}^{M}$ collect the rigid velocities
and the forces and torques on the $M$ colloids.
The geometric matrix $\bK$ transforms the rigid body velocities into surface velocities,
and its transpose integrates the single layer potential to give the external force and torque on the bodies, Eqs.\ \eqref{eq:balanceF}-\eqref{eq:balanceT}.
The velocity solution of \eqref{eq:linear} is
\eqn{
  % \bU = \bN \bF + \wtil{\bN} \pare{\bu_s + \bu_{\text{th}}},
 \bU = \bN \bF - \wtil{\bN} \bu_s - \wtil{\bN}_{\text{th}} \bu_{\text{th}},
}
where $\bN = \corchete{\bK^T \bM^{-1} \bK}^{-1}$ is the colloidal mobility matrix, and $\wtil{\bN} = \bN \bK^T \bM^{-1} \pare{\fr{1}{2}\bI + \bD_{St}}$ and 
 $\wtil{\bN}_{\text{th}} = \bN \bK^T \bM^{-1}$ are the slip mobility matrices that relate the particle velocities with their surface slip flows and thermal fluctuations respectively.

We can show that the covariance of the rigid body velocities generated by the stochastic slip obeys the fluctuation dissipation balance.
Setting $\bF=\bzero$ and $\bu_s=\bzero$ and using the definition of the mobilities $\bN$ and $\wtil{\bN}_{\text{th}}$ and the thermal noise, Eq.\ \eqref{eq:u_th},
one can compute the velocity covariance with a pure linear algebra derivation
\eqn{
\label{eq:fluctu-diss}
  \avg{\bU \bU^T} &= \wtil{\bN}_{\text{th}} \avg{\bu_{\text{th}} \bu_{\text{th}}^T} \wtil{\bN}_{\text{th}}^T  \nonumber \\
    &= (\bN \bK^T \bM^{-1})  \sqrt{2\kt} \bM^{1/2} \avg{ \bZ \bZ^T } \bM^{1/2}  \sqrt{2\kt} (\bN \bK^T \bM^{-1})^T  \nonumber  \\
    &= 2\kt \bN \bK^T \bM^{-1}  \bM  \bM^{-1} \bK \bN  \,\delta(t - t') \nonumber  \\
    &= 2\kt \bN (\bK \bM^{-1} \bK^T) \bN  \,\delta(t - t')  \nonumber  \\
  &= 2\kt \bN \,\delta(t - t').
}
Eq.\ \eqref{eq:fluctu-diss} shows that the thermal noise is balanced by the viscous dissipation as required by the fluctuation dissipation balance,
thus the slip \eqref{eq:u_th} generates Brownian velocities consistent with the Stokes equation \cite{Delong2014a, Sprinkle2019}. 
Once the velocities have been computed the orientation of the colloids, described by the unit quaternions $\btheta_m$, and their positions, $\bq_m$, can be integrated in time
\eqn{
  \label{eq:dynamics}
  \dd \bx = \bU \bullet \dd t, 
  % \{\dd \bq_m, \dd \btheta_m \}_{m=1}^M = \{\bu_m, \bomega_m \}_{m=1}^M \bullet \dd t
}
where $\bx = \{\bq_m, \btheta_m\}_{m=1}^M$.
We have introduced the abstract product notation $\bullet$ to emphasize that the numerical integration should conserve the unit norm of the quaternions to correctly represent orientations
and that, in the case of Brownian simulations, \eqref{eq:dynamics} is a multiplicative stochastic equation that should be integrated numerically with a stochastic integrator.
Both issues are discussed in detail in Refs.\ \cite{Sprinkle2017, Westwood2022}.
%To integrate the equations of motion  we can use a simple Euler or Midpoint scheme for deterministic simulations, i.e.\ $\bu_{\text{th}}=\bzero$, or a stochastic integrator for Brownian simulations \cite{Sprinkle2017, Westwood2022}.

To solve the linear system \eqref{eq:linear} we do not need to form the mobility matrices $\bN$ and $\wtil{\bN}$, whose computational cost would scales as $\mc{O}\pare{N_b^3}$.
Instead, we use a block-diagonal preconditioned GMRES which converges in a number of iterations independent of the number of colloids \cite{Usabiaga2016}.
Thus, all the methods that we employ to solve the hydrochemical problem with thermal fluctuations are matrix free and can use fast methods
to compute the action of the Laplace and Stokes kernels, such as the Fast Multipole Method \cite{Yan2020}.
This approach allows to simulate system with a large number of colloids as we will show in Sec.\ \ref{sec:Simulations}.

One can see that we use essentially the same formalism to discretize both the Laplace and the Stokes equations. 
The main difference is that to solve the Stokes equation we use a regularized version of its Green's function. 
Note that for non-overlapping nodes the RPY approximation, Eq.\ \eqref{eq:RPY}, obeys for any kernel $\bG(\br, \br')$
\eqn{
  \label{eq:RPY_no_overlapping}
  \bM_{ij} = 
  \pare{\bI + \fr{a^2}{6} \bna^2_{\br}} \pare{\bI + \fr{a^2}{6} \bna^2_{\br'}} \left. \bG(\br, \br') \right\vert_{\br'=\br_j}^{\br=\br_i}.
}
For the Laplace kernels this regularization would not introduce any correction because,
as the Laplace equation is harmonic, the Laplacian terms in \eqref{eq:RPY_no_overlapping} would vanish. 
The same argument applies to the derivatives of the Laplace kernel. 
Moreover, the positive definiteness of the mobility matrix, $\bM$, is fundamental to generate the stochastic slip,
while for the Laplace equation it is not such a fundamental property.

\subsection{Convergence of the iterative solvers}
\label{sec:convergence}

\begin{figure}[!htb]
  \centering
  \includegraphics[width=0.85 \textwidth]{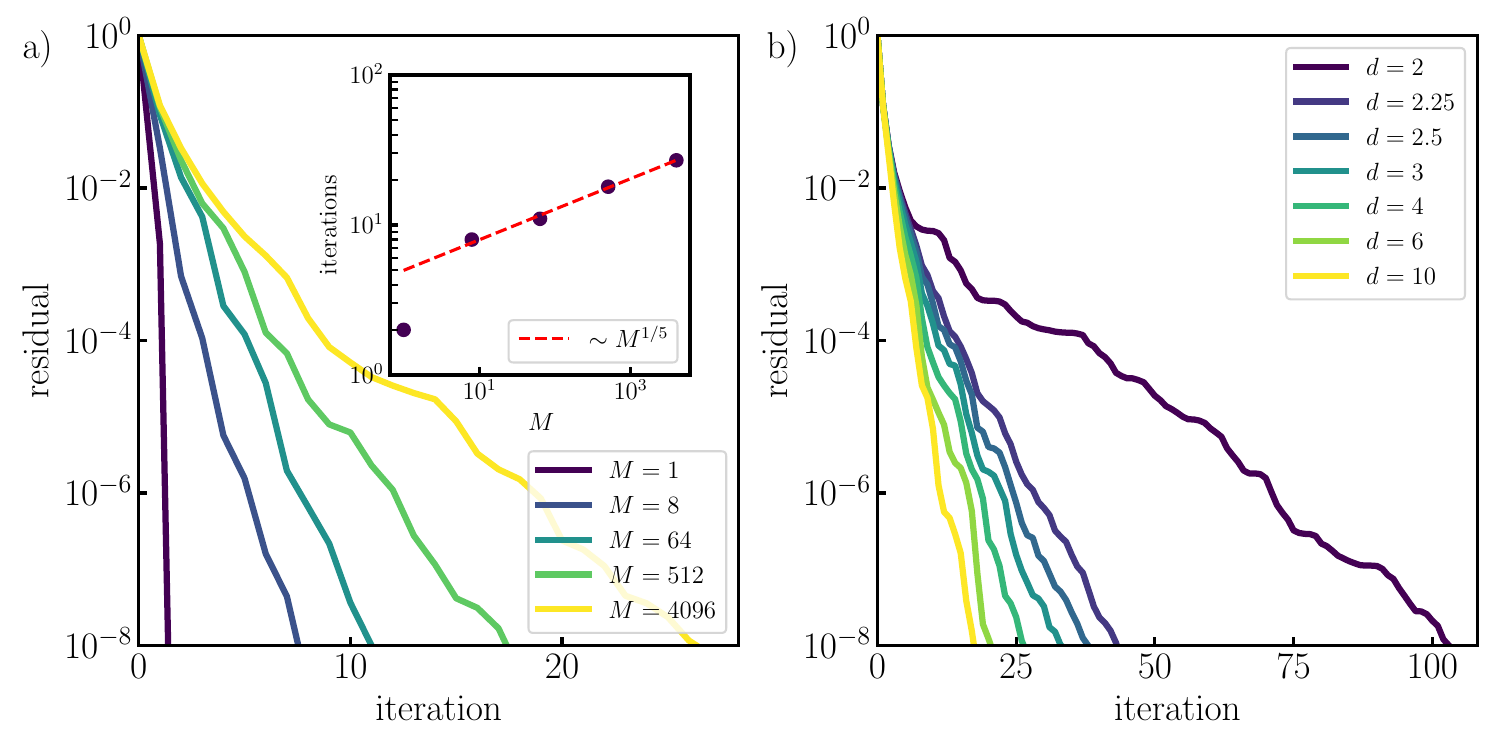}
  \caption{Linear solver convergence to solve the Laplace problem for reacting shells ($k=1, \mu = 1$ and $\alpha=0$) forming a simple cubic lattice in a constant background field $c_{\infty}=1$.
    Panel a) shows the convergence versus the number of shells, $M$, for lattice spacing $d=4$ shell radius.  
    The main figure shows the GMRES residual while the inset shows the number o iterations to attain a tolerance of $10^{-8}$ versus the number of shells.
    Even without preconditioner the number of iterations grows slowly ($\sim M^{1/5}$) with the number of shells.
    Panel b) shows the convergence versus the lattice spacing, $d$, for a lattice with $M=4096$ shells.
    GMRES converges quickly except for touching spheres, $d=2$. 
  }
  \label{fig:gmres}
\end{figure}

The second layer formulation of the Laplace problem, Eq.\ \eqref{eq:complete_v2}, is well conditioned.
Thus, the linear system  \eqref{eq:complete_discrete} can be solved with an iterative solver, such as GMRES, without preconditioner even for large systems.
To verify how robust is the convergence, we solve the Laplace problem for a set of uniform reactive shells ($k=1, \mu = 1$) forming a simple cubic lattice,
immersed in a constant background concentration $c_{\infty}(\br) = 1$.
We discretize the shells, of radius $R=1$, with $42$ nodes and we vary the number of shells, $M$, and the distance between first neighbors $d$.
We show in Fig.\ \ref{fig:gmres}a the GMRES convergence for moderately separated shells ($d=4$) versus the system size.
We observe that even for large systems, $4096$ shells or $172032$ nodes, the linear solver converges in a moderate number of iterations.
In fact, the number of iterations grows weakly with the number of shells $\sim M^{1/5}$.
We show in Fig.\ \ref{fig:gmres}b the convergence for a large system, $M=4096$, and different interparticle distances.
Denser packings require more iteration as expected but, except for touching shells $d=2$, the solver converges to a residual  $< 10^{-8}$ in less than 50 iterations.

The linear system for the Stokes problem \eqref{eq:linear} is solved with a preconditioned GMRES algorithm that keeps the number of iterations independent of the number of colloids as shown and explained in Ref.\ \cite{Usabiaga2016}.

\section{Grid optimization}
\label{sec:Grid_opt}

The goal of this section is to optimize the surface grid in order to match the hydrodynamic response of an ideal particle to external forces and torques, $\bF$, and to slip surface flows, $\bu_s$, regardless of their origin.
While the use of the regularized Green's function \eqref{eq:RPY} simplifies the generation of the Brownian noise, Eq.\ \eqref{eq:u_th},
and makes the rigid multiblob method robust,  its error convergence rate is relatively slow with respect to the number of discretization nodes.
In order to evaluate the effect of the regularization in the continuous setting, i.e.\ prior to discretization, one may  consider the flow induced by a known force density on a body's surface, $\blambda$, with the regularized Green's function \eqref{eq:RPY}
\eqn{
  \bv(\bx) = \int \bM(\bx, \by) \blambda(\by) \,\dd S_y = \int \bG_{\text{St}}(\bx, \by) \blambda(\by) \,\dd S_y
  + \int \Delta \bM(\bx, \by) \blambda(\by) \,\dd S_y,
}
where $\Delta \bM(\bx, \by) = \bM(\bx, \by) - \bG_{\text{St}}(\bx, \by)$ is the difference between the RPY mobility \eqref{eq:RPY} and the Green's function of the
Stokes equation.
The first term in the right hand side is the exact flow predicted by the Stokes equation. Thus, even
before the discretization step, the method introduces a regularization error in the flow.
The \emph{correction} kernel scales with the node radius, $a$, as
\eqn{
  \Delta \bM(\bx, \by) \sim
  \begin{cases}
    a^2 & \text{for } |\bx - \by| > 2a, \\
    a & \text{for } |\bx - \by| \le 2a.
  \end{cases}
}
Accordingly, given that the discretization employs a number of blobs $N_b \sim 1 / a^2$, the convergence rate of the error scales as $\sim N_b^{-1/2}$ for self interactions,
i.e.\ for the self-mobility matrices of the bodies.

Despite this regularization error, it is possible to enhance the numerical accuracy for a given resolution through a grid optimization procedure, the details of which are outlined below.\\

In the absence of thermal fluctuations, the hydrodynamic response of the particle is given by the solution of the mobility problem:
\eqn{
\bU  =  \bN\bF - \wtil{\bN}\bu_s
}
where $\wtil{\bN}=\bN\bK^T\bM^{-1}(\fr{1}{2}\bI + \bD_{\tex{St}})$ is the slip mobility matrix and $\bD_{\tex{St}}$ is the Stokes second layer operator, both introduced in Section \ref{sec:disc_Stokes}. The matrix $\wtil{\bN}$ maps the slip velocities to the forces applied on the surface nodes, through the node resistance operator $\bM^{-1}$, then computes their resultant on each rigid body, through the operator $\bK^T$, and multiply the resulting forces and torques by the body mobility matrix $\bN$ to obtain the slip-induced body velocities in the suspension. 
 As shown in \cite{Swan2011}, a continuous version of $\wtil{\bN}$, which generalizes the work of Stone and Samuel \cite{StoneSamuel1996}, can be derived using the Lorentz reciprocal theorem. 

% \Blaise{Is it possible to show that $\wtil{\bN}$ is the average of the slip flow for a spherical particle ? cf. eq.\\ (4)-(5) of Stone Samuel \cite{StoneSamuel1996}. Yes, through the pseudo inverse of $\bK$!
% But first I need to prove that $\bu^s$ belongs to the column space of $\bK$ for a spherical (and maybe other) shape  (so that $\bK \bK^+\bu^s = \bu^s$). Indeed if $\bu^s$ belongs to the column space of $\bK$ then $$\bU = \bN\bK^T\bM^{-1}\bu^s = \bN\bK^T\bM^{-1}\bK\bK^+\bu^s = \bK^+\bu^s$$ because $\bK\bK^+$ is just a projector on the column space of $\bK$.

% For a spherical particle, 
% $$\bK^+ = \left[\begin{array}{c}
% \fr{1}{N_b} \bK^T_{\bI}\\
% \fr{1}{N_b}\fr{3}{4 R^2} \bK^T_{\br^{\times}}\\
% \end{array}\right],$$
% where $\bK^T_{\bI}$ corresponds to the first three rows of $\bK^T$ and $\bK^T_{\br^{\times}}$ to its last three rows.  Thus $\bK^+$ corresponds exactly to the averaging operators derived by Stone and Samuel eq.\\ (4)-(5).

% EVEN BETTER: Swan \cite{Swan2011} generalized Stone's formula for both rotation and translation for bodies of arbitrary shape in arbitrary geometries.
% }

For an isolated body discretized with $N_b$ nodes,  $\bN$ is a $6\times 6$ matrix and $\wtil{\bN}$ is a  $6\times 3N_b$ matrix.
Let $\bN_{ref}$ and $\wtil{\bN}_{ref}$ be the reference mobilities to be matched for a given body, which can be known analytically or calculated with a well-resolved discretization.
$\bN$ and $\bN_{ref}$ have the same $6 \times 6$ shape  regardless of the solution method used to compute them.
However, $\wtil{\bN}$ and $\wtil{\bN}_{ref}$ have different sizes since their number of columns depends on the grid resolution, $N_b$, regardless of the numerical method used to compute the reference solution.

In their recent work, Broms \textit{et al.} \cite{Broms2023} proposed a systematic approach to match the mobility matrix $\bN_{ref}$ for spherical and rod-like passive particles. They did so by optimizing the grid, and more specifically the geometric dimensions of the grid surface (length, radius) and the node size $a$ (or equivalently the node spacing). 

Here we extend their framework to include the slip-mobility matrix $\wtil{\bN}_{ref}$ in the optimization problem. 
In order to generalize the method to arbitrary shapes, we consider only two parameters to be optimized:  the size scale of the body, a scalar that we denote  as $S$ (e.g. $S=R_g/R$ for a sphere of radius $R$, where $R_g$ is the geometric radius of the surface grid)  and the radius of the nodes $a$. If $S = 1$, then the geometric surface of the grid, where the nodes are  located, coincides with the surface of the ideal particle. If $S<1$ ($S>1$) the geometric surface is located in the interior (exterior) of the ideal surface.

\subsection{Singular value decomposition of the slip mobility matrix}
\label{sec:SVD}
Since $\wtil{\bN}$ and $\wtil{\bN}_{ref}$ do not have the same size (usually $N_{b,ref}\gg N_b$), one way to compare these two matrices is through their singular values decomposition (SVD). Keeping only its compact form, the SVD decomposition of $\wtil{\bN}$ is 
\eqn{
\wtil{\bN} = \bW\bSigma \bV^T
}
where the $6\times 6$ matrix $\bSigma$ contains the $6$ singular values along the  diagonal
\eqn{
\bSigma = \left[\begin{array}{ccc}
\sigma_1 & 0 & 0 \\
0 & \ddots & 0  \\
0 & 0 &\sigma_6 \\
\end{array}\right].
}
As shown in \ref{app:scalingSV}, the singular values of the slip mobility matrix scale as $\sim N_b^{1/2}$. \\
In order to separate slip-translation ($US$) and slip-rotation ($\Omega S$) couplings,  we decompose $\wtil{\bN}$  into two $3\times 3N_b$ blocks, each of which are decomposed with SVD: 
\eqn{
\wtil{\bN} = \left[\begin{array}{c}
\wtil{\bN}^{US}\\
\wtil{\bN}^{\Omega S}\\
\end{array}\right] = \left[\begin{array}{c}
\bW^{US}\bSigma^{US}(\bV^{US})^T\\
\bW^{\Omega S}\bSigma^{\Omega S}(\bV^{\Omega S})^T\\
\end{array}\right].
}
Because these two blocks are independent, their  three singular values correspond to the singular values of the whole matrix.

To gain a bit more insight, we will focus on the $US$ block to provide a physical interpretation of the SVD decomposition.
The matrix $\bV^{US} = [\bv_1, \bv_2,  \bv_3]$ contains three orthonormal eigenmodes, i.e.\  unitary slip distributions on the particle surface. These distributions induce, through the action of $\wtil{\bN}^{US}$, translational motion along three orthonormal directions spanning $\mathbb{R}^3$ and contained in $\bW^{US} = [\bw_1, \bw_2,  \bw_3]$. The singular values $\{\sigma_{1},\sigma_{2}, \sigma_{3}\}$ provide the magnitude the resulting translational motion.
For simplicity, consider a spherical, isotropic, object, so that $\bSigma^{US} = \sigma \bI_3$. As a result the SVD can be written as 
$$\wtil{\bN}^{US} = \bW^{US}\sigma \bI_3 (\bV^{US})^T = \sigma (\bar{\bV}^{US})^T$$
where $\bar{\bV}^{US} = \bV^{US}(\bW^{US})^T = [\bar{\bv}_1,\bar{\bv}_2,\bar{\bv}_3]$ are the eigenmodes rotated in the frame of the translational motions $\bw$'s.
In this case, the hydrodynamic response to an arbitrary slip distribution $\bu_s$ is given by 
$$\wtil{\bN}^{US}\bu_s = \sigma \left[\begin{array}{c} \bar{\bv}_1\cdot \bu_s\\  
\bar{\bv}_2\cdot \bu_s\\
\bar{\bv}_3\cdot \bu_s
\end{array}\right].$$
Therefore, in order to match the hydrodynamic response of the reference grid, one must match both the singular value $\sigma$, and the projection of $\bu_s$ on the eigenmodes: $\bar{\bv}_i\cdot \bu_s$. Direct comparison of the  singular values between two different grids  is straightforward because their number does not depend on $N_b$ nor on the slip distribution $\bu_s$. On the other hand, the projected value depends both on the slip distribution $\bu_s$ and on the eigenmodes $\bar{\bv}$'s, and thus on the number of grid points. Therefore, in the following we decide to match the $\sigma$'s only. A few empirical tests have shown that the eigenmodes of two different discretizations of the same shape are similar and thus, if the $\sigma$'s match within a tight tolerance, then their hydrodynamic response to a given slip distribution should be comparable.

 % Note also that the mobility matrix $\bN$ appears in the slip-velocity coupling $\wtil{\bN} = \bN\bK^T\bM^{-1}$ so we could also define the error using only the singular values of the matrix $\bK^T\bM^{-1}$ instead and would get the same results \Blaise{to check, but I remember it was the case...}.\\

\subsection{Optimization procedure}
Following \cite{Broms2023}, our optimization procedure is based on the minimization of a cost function that measures the relative error between the  mobility matrices of the surface grid and a reference grid/solution. 
Since each block of the mobility matrices scale differently with the particle size (e.g. $1/R$ for force-translation couplings \textit{vs.}\ $1/R^3$ for torque-rotation couplings of a sphere), we define the errors between the reference and the computed mobility matrix block by block 
\eqn{
E_{UF} = \|\bN^{UF}_{ref} - \bN^{UF}\|/\|\bN^{UF}_{ref}\| \\
E_{\Omega F} = \|\bN^{\Omega F}_{ref} - \bN^{\Omega F}\|/\|\bN^{\Omega F}_{ref}\| \\
E_{UT} = \|\bN^{UT}_{ref} - \bN^{UT}\|/\|\bN^{UT}_{ref}\| \\
E_{\Omega T} = \|\bN^{\Omega T}_{ref} - \bN^{\Omega T}\|/\|\bN^{\Omega T}_{ref}\|}
as well as 
\eqn{
E_{US} = \|\bSigma^{US}_{ref} - \beta \bSigma^{US}\|/\|\bSigma^{US}_{ref}\| \\
E_{\Omega S} = \|\bSigma^{\Omega S}_{ref} - \beta \bSigma^{\Omega S}\|/\|\bSigma^{\Omega S}_{ref}\| \\
}
where $\beta = (N_b/N_{b,ref})^{1/2} \sim \|\bSigma_{ref}\|/\|\bSigma\|$ is the scaling factor between the singular values of the reference and discrete grids (see \ref{app:scalingSV}). The matrix norm chosen in this work is the Frobenius norm, but the choice of a specific norm does not affect significantly the outcome of the optimization.

From these relative errors we define the objective function $f$  to  minimize in order to find the optimal particle size $S$ and node radius $a$ as 
\eqn{
f(S,a) = w_{UF} E_{UF} + w_{\Omega F}E_{\Omega F} + w_{UT}E_{UT} + w_{\Omega F}E_{\Omega T} + w_{US}E_{US} + w_{\Omega S}E_{\Omega S},
\label{eq:cost_func}
}
where $\{w_{UF},\cdots,w_{\Omega S}\}$ are weights that can be adjusted depending on the types of hydrodynamic response that are privileged with respect to the others. Unless specified otherwise, the weights are taken to be equal hereafter ($w_{UF}=\cdots=w_{\Omega S}=1$).  We also considered another cost function using the max of the errors  as in \cite{Broms2023} but found \eqref{eq:cost_func} to be more robust and that performs well for our purposes (see  \ref{app:minmax}). 
% and 
% \eqn{
% f_2(S,a) = \max\{E_{UF}, E_{\Omega F}, E_{UT},  E_{\Omega T}, E_{US}, E_{\Omega S}\}.
% }
The minimization problem to be solved is 
\eqn{
& \underset{S,a}{\min}f , \label{eq:opt_pb}  \\
\text{s.t.} \, &\, 0<a<L/2, \nonumber \\
  &\, 1/2<S<2 \nonumber
 }
where $L$ is a typical length scale of the particle (e.g.\ $L = R$ for a sphere). The bounds added in \eqref{eq:opt_pb} are used to avoid unphysical solutions and to reduce the size of the domain explored in parameter space.\\ 
The problem is solved with a differential evolution algorithm \cite{Storn1997}, which, due to its stochastic nature and to the small dimension of the parameter space, systemically converges  to a global minimum (while we found that gradient-based methods were much more sensitive to the initial guess and sometimes stopped at local minima).

\section{Validations}
\label{sec:Validations}
In this section, we first consider the motion of slip-driven particles to evaluate the performance of the grid optimization procedure described in Section \ref{sec:Grid_opt}. Then, we validate our Laplace and Stokes solvers with canonical cases against reference solutions from the literature. We show that, thanks to the double layer formulation of the Stokes problem and to the optimized grids, we achieve a good agreement with exact solutions with only a few nodes on the particle surface. 

\subsection{Swimmers with a prescribed slip}
\label{sec:valid_slip}
\subsubsection{Spherical squirmer}
\label{sec:squirm}
The spherical squirmer model consists of a spherical particle with a prescribed surface slip velocity, i.e.\ that does not depend on concentration or any other quantity, usually decomposed into series of spherical harmonics \cite{Pak2014}.  It has been extensively used to model the self-propulsion and flow generated by  ciliated microorganisms  \cite{Lighthill1952,Blake1971,Pedley2016}. 

Here we consider a spherical squirmer of radius $R$ in an unbounded domain with an axisymmetric slip distribution on its surface. The exact mobility coefficients of the sphere are given by the Stokes drag ($N^{UF}_{\text{th}} = 1/(6\pi \eta R)$ and $N^{\Omega T}_{\text{th}} = 1/(8\pi \eta R^3)$) while the exact slip-velocity couplings are given by surface averages of the slip flow, see Eqs.\ (4) and (6) in \cite{StoneSamuel1996}.

With the multiblob method,  the sphere surface is discretized with nodes located on geodesic grids \cite{Swan2016}. In the original works of \cite{Swan2016} and\cite{Usabiaga2016}, the node radius $a$ is chosen so that neighboring nodes touch each other ($a/s=0.5$, where $s$ is the nearest-neighbor distance) and the geometric radius of the grid $R_g$ is chosen in order to match the Stokes drag  of an ideal sphere with radius $R$ in an unbounded domain (i.e.\ $N^{UF}_{\text{th}} = 1/(6\pi \eta R)$). We denote this discretization as the ``original grid''. We also define the ``optimized grid'', parametrized by the doublet $(S = R_g/R ,a)$, as the solution of the minimization problem \eqref{eq:opt_pb} that includes \textit{all} the mobility coefficients, as in \cite{Broms2023},  \textit{and} slip-velocity couplings (which is the new part from our contribution).
The reference mobility and slip-mobility matrices, $\bN_{ref}$ and $\wtil{\bN}_{ref}$, are obtained from a highly resolved sphere with $N_{b,ref} = 10242$ nodes. The reference mobility coefficients are close to the analytical ones (relative error is $3\times 10^{-5}$ for the $UF$-coefficient , and $2\times 10^{-4}$ for the $\Omega T$-coefficient), and  the propulsion speed $\bU_{ref}$ due to $\bu_s$ has a relative error of $8\times 10^{-3}$ with respect to $\bU_{\text{th}}$. 

\begin{figure}[!htb]
  \centering
  \subfigure[]{  \includegraphics[width=0.43\textwidth]{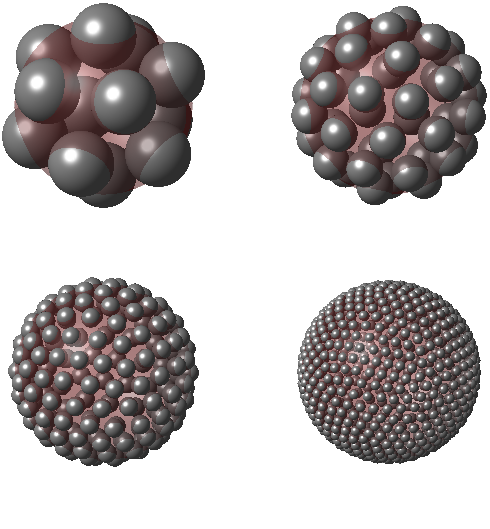} }
  \hfill
  \subfigure[]{\includegraphics[width=0.5\textwidth]{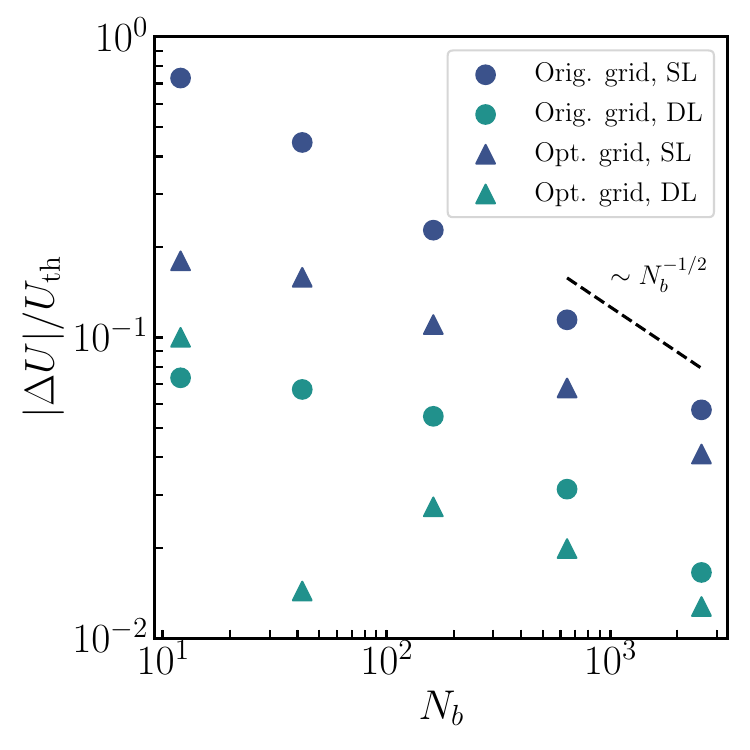}}
  \caption{a) Optimized discretization of the sphere surface with the double layer formulation (DL) for $N_b = 12, 42, 162$ and $642$ nodes respectively (see last column of Table \ref{tab:opt_res}). The ideal sphere ($R=1$) is shown in light red. b) Relative error of the propulsion speed of a squirmer as a function of the grid resolution $N_b$.}
  \label{fig:squirmer_speed_optimized}
\end{figure}

We consider two different discrete formulations of the integral equation \eqref{eq:slip_integral}: the double layer (DL) formulation \eqref{eq:RBM_slip_discrete}, and the single layer (SL) formulation, where  the expression  $\pare{\fr{1}{2}\bI + \bD_{\tex{St}}} \bu_s$ in \eqref{eq:RBM_slip_discrete} is replaced by  $\bu_s$ (see Section \ref{sec:integral_Stokes}).
Table \ref{tab:opt_res} shows the original and optimized grids using the SL and DL formulations, together with the optimized grid in the absence of slip, where only the mobility matrix $\bN$ is matched. As expected, our optimized grid in the absence of slip matches exactly the one from  Broms \textit{et al.} (see Table 2 of \cite{Broms2023}). We first notice that adding the slip mobility matrix in the cost function changes  both the size scale $S$ but also the relative node size $a/s$ compared to the mobility problem without slip. However these changes differ between the SL and DL formulation. In the SL case, the optimized grid is very close to the geometric surface of the sphere ($S > 0.99$) and the relative node size increase with $N_b$ but remains well below contact ($a/s<0.35$). In the DL case, the optimized grid, shown in Fig.\ \ref{fig:squirmer_speed_optimized}a, has a smaller radius ($S<0.99$) and the nodes remain close to contact ($a/s\geq 0.41$) for all resolutions, which reduces the leakage of the fluid through the gaps between the nodes compared to the SL optimized grids. 
Indeed,  when the flow is squeezed against the particle surface, e.g.\ due to the presence of an other particle, it will leak more through the SL grid compared to the DL grid.
Leakage across discretized surfaces has also been observed in the context of Immersed Boundaries \cite{Griffith2012,Kallemov2016,Bao2017} and Regularized Stokeslet \cite{Hoffmann2017,Cortez2018,Chisholm2022} when the nodes (or Regularized Stokeslets)  are too far apart from each other.

\begin{table}[htb!]
  \begin{center}
    \begin{tabular}{ c |c c | c c | c c | c c}
      & \multicolumn{2}{c|}{Orig.\ grid} &  \multicolumn{2}{c|}{Opt.\ grid w/o slip}  & \multicolumn{2}{c|}{Opt.\ grid SL}  & \multicolumn{2}{c}{Opt.\  grid DL} \\
      \hline
      $N_b$ & $S$  & $a/s$ & $S$  & $a/s$ & $S$  & $a/s$ & $S$  & $a/s$\\
      \hline
      12& 0.792 & 0.5 & 0.936 & 0.291 & 0.991 & 0.231 & 0.858 & 0.412 \\
      42& 0.891 &  0.5 & 0.959 & 0.311 &  0.992 &  0.238 & 0.929 &  0.410\\
      162& 0.950 & 0.5 & 0.974 & 0.345  & 0.992 & 0.254 &  0.965 & 0.413\\
      642&  0.977 & 0.5 & 0.984 & 0.388 &  0.993 & 0.284 &  0.982 & 0.424\\
      2562& 0.989 & 0.5 & 0.991 &0.431 & 0.994 & 0.346 & 0.990 & 0.449 \\
    \end{tabular}
  \caption{ Original and optimized grid parameters for a sphere with radius $R = 1$ in an unbounded domain.}
  \label{tab:opt_res}    
  \end{center}
\end{table}

We now evaluate the hydrodynamic response of the sphere to a surface slip given, in the frame of the particle, by $\bu_s = -\sin \theta \be_{\theta}$, where $\theta \in [0,\pi]$ is the angle between the position on the squirmer surface and its axis of symmetry $\be_z$. It is known from \cite{Lighthill1952} that the resulting swimming speed is $\bU_{\text{th}} = 2/3 \be_z$.

We compare the performance of each grid using both  the single layer (SL)  and the double-layer (DL) formulation. As explained in Section  \ref{sec:integral_Stokes}, even though the double layer operator can theoretically be eliminated from the boundary integral formulation with slip flows, it has been shown that its presence in the discretized version can drastically reduce numerical errors \cite{Smith2021}. 
Figure \ref{fig:squirmer_speed_optimized}b compares the propulsion speed from the different grids against the exact solution. Due to the regularization error (see Section \ref{sec:Grid_opt}), the  grids converge slowly,  as $\sim N_b^{-1/2}$, to the exact solution. The error of the original grid with SL (blue disks) is quite large  for low and intermediate resolutions ($\geq 22\%$ for $N_b \leq 162$). This error is significantly reduced with the optimized grid (blue triangles), but remains above  $11\%$ for $N_b \leq 162$. The  DL formulation (green disks) drastically decreases the error on the original grid: it always remains below $10 \%$ for all resolutions and  below $6\%$ above $N_b\geq 42$. Grid optimization further improves the error which drops below $3\%$ for $N_b\geq 42$. 

Altogether these results confirm that the DL formulation \eqref{eq:slip_discrete} is crucial to properly account for surface flows with a low resolution, hence this is the formulation adopted hereafter in this work. On top of that, our grid optimization technique further decreases the error by a factor 2, which permits to simulate slip-driven particles with a coarse grid ($N_b = 42$) with a 2-3 digits accuracy.

\subsubsection{Slip-driven smooth rod}
\label{sec:smooth-cyl-valid}
\begin{figure}[!htb]
  \centering
  \includegraphics[width=0.9\textwidth]{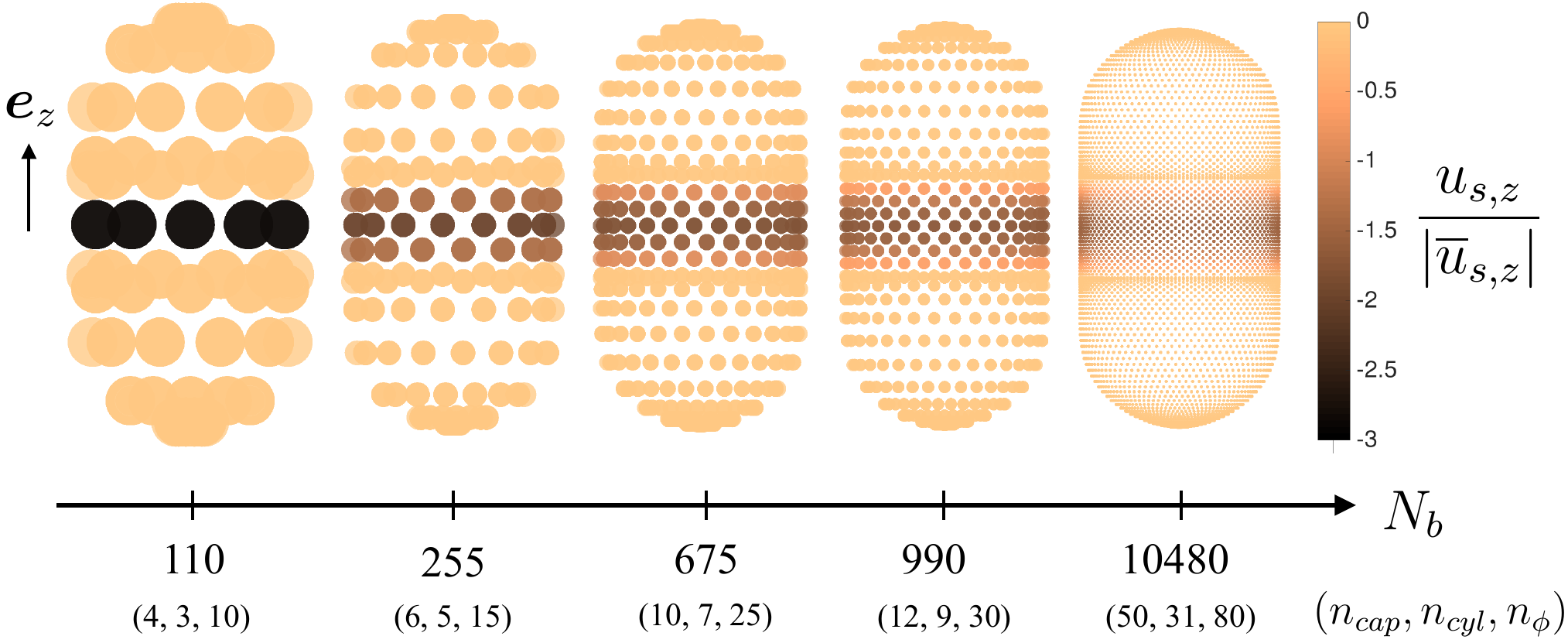}
  \caption{Discretization of the smooth rod with different resolutions (original grid: $S=1, a/s =0.5$). Colorbar: prescribed slip at the node centers renormalized by the mean value of the continuous distribution $\overline{u}_{s,z} = 2/\pi$. }
  \label{fig:smooth_rod_disc}
\end{figure}

In this section we consider the motion of a rod-like particle with aspect ratio $L/R = 4$ (length $L = 2 \mu$m, radius $R = 0.5 \mu$m) due to a prescribed slip velocity on its surface. Following  \cite{Broms2023} and \cite{Bagge2021} we discretize the rod surface with a  quadrature that ensures the continuity of the normal vector $\bn$. The  discretization of the rod is divided into three parts: two caps, each discretized with $n_{cap}$ Gauss-Legendre nodes along the curvilinear length, connected by a central cylinder, discretized with $n_{cyl}$ equally spaced nodes along the axial direction. Both the caps and the cylinder are discretized with $n_\phi$ nodes along the azimuthal direction. The total number of nodes per cylinder is therefore $N_b = (2n_{cap} + n_{cyl})n_{\phi}$. The cross-sections are shifted so that every second layer is aligned. An ``original" grid is chosen naively by placing the nodes on the actual surface of the rod ($S=1$), and the node radius is chosen so that the distance between two adjacent layers of the central cylinder, $s$, is equal to one node diameter ($a/s = 0.5$). Figure \ref{fig:smooth_rod_disc} shows the original grid for five different resolutions. Due to its axisymmetry, the particle has only four independent diagonal mobility coefficients in an unbounded domain: $N^{UF}_{\parallel}, N^{UF}_{\perp}, N^{\Omega T}_{\parallel}, N^{\Omega T}_{\perp}$, where $\parallel$ and $\perp$ are the direction parallel and perpendicular to the axis of symmetry of the rod respectively.  
Note that, because of the shift between contiguous cross-sections, $n_{cyl}$ must be an odd number in order to preserve the fore-aft symmetry of the particles and to avoid spurious nonzero off-diagonal coefficients in the mobility matrix, which was overlooked by \cite{Broms2023}.

\begin{figure}[!htb]
  \centering
  \includegraphics[width=0.49\textwidth]{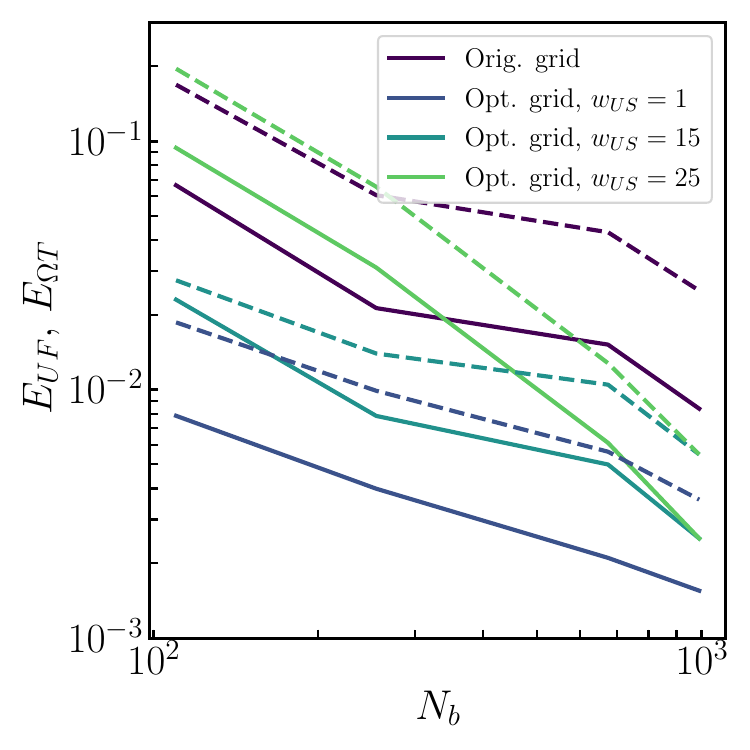}
  \includegraphics[width=0.49\textwidth]{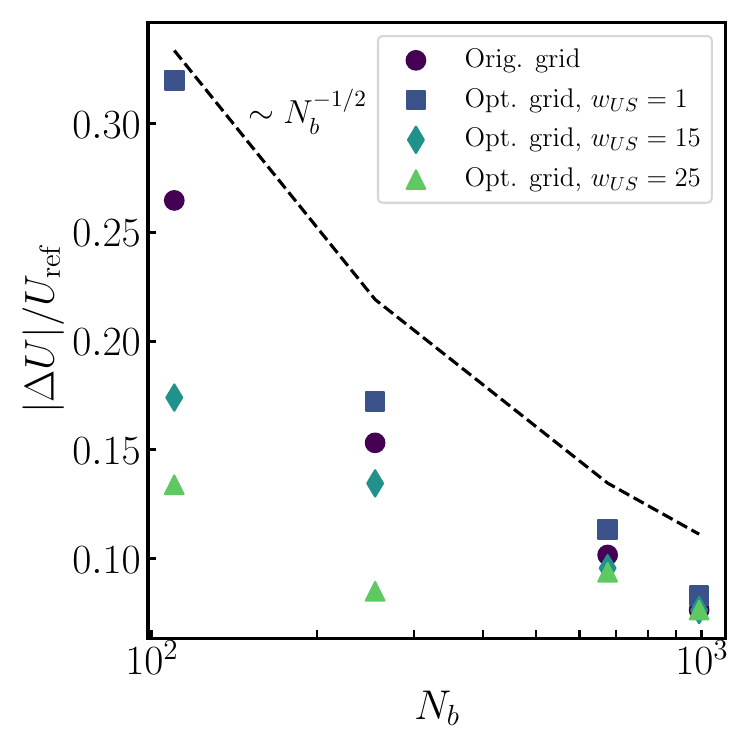}
  \caption{\textbf{(Left)} Relative error of the rod mobility coefficients as a function of the grid resolution $N_b$ for various values of the slip-translation weight $w_{US}$. Solid line: $UF$ couplings. Dashed line: $\Omega T$ couplings. \textbf{(Right)} Relative error of the propulsion speed of a slip-driven rod as a function of the grid resolution $N_b$ for various values of the slip-translation weight $w_{US}$.}
  \label{fig:result_optim_smooth_rod}
\end{figure}

The reference grid used for the optimization process is the original discretization ($S=1$, $a/s=0.5$) with $N_b = 10480$ nodes ($n_{cyl} = 30$, $n_{cap} = 50$, $n_\phi = 80$). Its mobility coefficients match the BEM solution from \cite{Broms2023} (Table A.12) within less than $0.1\%$ for the UF coefficients and less than $0.7\%$ for the $\Omega T$ coefficients.
Table \ref{tab:opt_res_cyl} provides the original and optimal grids for various values of the weight of the slip-translation coupling $w_{US}$ in the cost function \eqref{eq:cost_func} ($w_{US} = 1$ is the default value and corresponds to an identical weight for all error contributions). As for the spherical case, the nodes on the optimal grid are located in the interior of the actual particle ($S<1$).  We also observe that, for a given resolution $N_b$, the relative node size $a/s$ increases with $w_{US}$ while the size scale $S$ varies nonmonotonically. 
%Finally, the finest optimal grid  is identical between $w_{US} = 15$ and $w_{US} = 25$ \Blaise{interpretation ?}
The resulting error on the mobility coefficients is shown in Figure \ref{fig:result_optim_smooth_rod}(left). When the weight on the slip-translation coupling is identical to the others ($w_{US} = 1$), the error on the mobility is decreased by one order of magnitude with respect to the original grid (from $O(10^{-1}-10^{-2})$ to $O(10^{-2}-10^{-3})$). However, as the weight increases, the error on the mobility increases as well, due to its smaller weight in the cost function. \\

\begin{table}[htb!]
\begin{center}
\begin{tabular}{ c |c c | c c | c c | c c} 
& \multicolumn{2}{c|}{Orig.\ grid} &  \multicolumn{2}{c|}{Opt.\ grid, $w_{US} = 1$}  & \multicolumn{2}{c|}{Opt.\ grid, $w_{US} = 15$}  & \multicolumn{2}{c}{Opt.\  grid, $w_{US} = 25$} \\
 \hline
 $N_b$ & $S$  & $a/s$ & $S$  & $a/s$ & $S$  & $a/s$ & $S$  & $a/s$\\
 \hline
110& 1 & 0.5 & 0.931 & 0.502 & 0.897 & 0.694 & 0.928 & 0.868  \\
255& 1 &  0.5 & 0.964 & 0.616 &  0.948 & 0.799  & 0.954 &  0.965\\
675& 1 & 0.5 & 0.980 & 0.570  & 0.972 & 0.744 &0.972  & 0.758\\
990&  1 & 0.5 & 0.987 & 0.567 & 0.984 & 0.660 &   0.984 & 0.660 \\
\end{tabular}
\caption{Original and optimized grid parameters for a smooth rod with aspect ratio  $L/R = 4$ in an unbounded domain.}
\label{tab:opt_res_cyl}
\end{center}
\end{table}

We now evaluate the hydrodynamic response of these discretized rods due to a surface slip velocity distributed over the central cylinder, given, in the frame of the particle, by 
\eqn{
\bu_s = u_{s,z}(z) \be_z =  -\cos \pare{\frac{\pi z}{L_c}} \be_z,\, \, \,\, z\in[-L_c/2, L_c/2],
}
where $\be_z$ is the axis of symmetry of the rod, $z=0$ corresponds to the midpoint of the central cylinder  and $L_c$ its length. In the discrete setting, the slip velocity of each node belonging to the cylinder surface is multiplied by a factor $A = \fr{\overline{u}_{s,z}}{\sum_i u_{s,z}(z_i)} $ to ensure that the slip distribution has the same mean ($\overline{u}_{s,z} = 2/\pi$) as the continuous one regardless of the  resolution (see Figure \ref{fig:smooth_rod_disc}).  Figure \ref{fig:result_optim_smooth_rod}(right) shows the relative error of the propulsion speed of the discrete rods with respect to the reference solution as a function of the grid resolution for $w_{US} = 1,\,15$ and $25$. For $w_{US} = 1$, the speed error is higher than the one of the original grid. However, as $w_{US}$ increases, the error decreases and eventually drops by a factor 2 compared to the original grid for the two coarsest grids ($N_b = 110$ and $255$).  Similarly to the spherical squirmer, the error converges as $N_b^{-1/2}$ and the effect of the optimization is less pronounced when the grid resolution increases.  

Altogether these results show that, for nonspherical geometries, it becomes more challenging to match both the mobility coefficients and the response to active surface slip simultaneously. Indeed, the mobility errors and speed errors have opposite trends with respect to the weight $w_{US}$. In the present case, the optimal grid obtained for $w_{US} = 15$ seems to be a good compromise as the mobility errors drop by a factor $\approx 3$ to 7 and the speed error decreases by $35 \%$ for the coarsest grid. 
Of course, these results are specific to the particle geometry and slip distribution presented here. They solely intend to show that by tuning the weights in the cost functions \eqref{eq:cost_func} one can find a compromise between the hydrodynamic response to forces/torques and to surface slip.

We also would like to emphasize that, in this work, we have chosen to write a simple and generic optimization problem which could be improved in many ways by taking into account the specificities of the particles considered. For instance, one could further improve the results by optimizing more geometric parameters such as the geometric length $L_g$ and radius $R_g$ of the grid, instead of a single size scale parameter $S$, and/or use different grids for the $UF$ and $\Omega T$ couplings, as in \cite{Broms2023}, but also for the $US$ and $\Omega S$ couplings.

\subsection{Inert particle in a linear concentration field}
\label{sec:passive_linear_field}

\begin{figure}[!htb]
  \centering
  \includegraphics[width=0.85\textwidth]{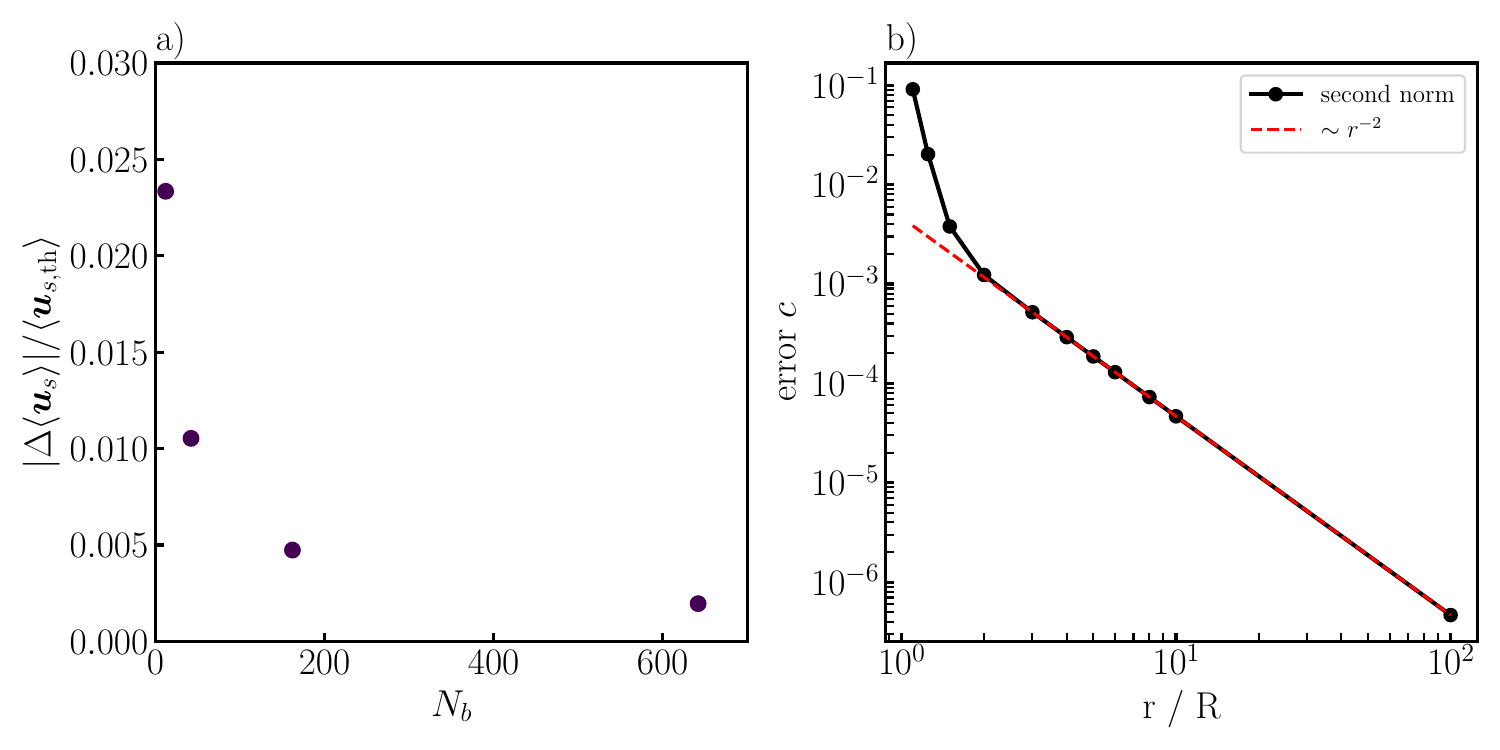}
  \caption{{\bf (a)} Relative mean slip error for a passive particle immersed in a external linear field as a function of the number of nodes.
    {\bf (b)}
    Concentration field error computed on a shell at a distance $r$ from the center of the colloid of radius $R$ discretized with $N_b = 42$ nodes.
    The error decays with the square of the distance for both the second and the infinite norm.
  }
  \label{fig:passive_polarity_slip}
\end{figure}

To validate the Laplace solver we study a single inert spherical colloid immersed in a linear concentration field 
$\bna c_{\infty}(\bx) = \bs{e}_z$. 
Due to the no-flux boundary conditions on its surface, the particle perturbs the external concentration field. 
To characterize the quality of the solution, we measure the average slip velocity induced by this disturbance which, for a constant surface mobility $\mu(\br)=1$,
is just the surface average  of the tangential concentration  gradient: $\avg{\bu_s}=\avg{\bna_{\parallel}c}$. 
The slip is azimuthal by symmetry and its average magnitude is $\avg{u_{s,th}}=1$ \cite{Rojas-Perez2021}. 
In this test we use the optimized resolutions given in the last column of Table \ref{tab:opt_res}.
Fig.\ \ref{fig:passive_polarity_slip}a shows the relative error in the average slip for colloids discretized with a different number of nodes $N_b$. 
We observe that the error is around $1\%$ with a discretization of 42 nodes, and below for higher resolutions. 
Although for us the concentration is an intermediate variable, as we are mostly interested in the velocity of the colloids,
we show in Fig.\ \ref{fig:passive_polarity_slip}b the absolute error on the concentration field computed on a shell of radius $r$ centered in the colloid.
We see that the error decays with the square of the distance and that it is of small magnitude, $O(10^{-3})$,  
%$\text{error } c \sim 10^{-3}$,
one radius away from the colloidal surface
for a colloid discretized with $N_b=42$ nodes.

\subsection{Active particles}
\subsubsection{Active particle with finite Damkholer number}
\label{sec:damkholer}

For reactive particles that consume reactant it is useful to define the dimensionless Damköhler number, $\text{Da} = k R / D$, see Section \ref{sec:eqns}.
Here we verify that our approach works for arbitrarily Da numbers by simulating a single Janus particle with an inert hemisphere and an active one
and by comparing our results against those of  Michelin and Lauga \cite{MichelinLauga2014}.
The surface mobility is set to $\mu(\br)=1$ over the whole sphere; 
the background concentration is set to zero, $c_{\infty}=0$, and
the emitting fluxes and reaction rates are set to $\alpha=-1$ and $k= \text{Da} D/R$ on the active hemisphere and $\alpha=k=0$ on the passive one.

\begin{figure}[!htb]
  \centering
\includegraphics[width=0.5\textwidth]{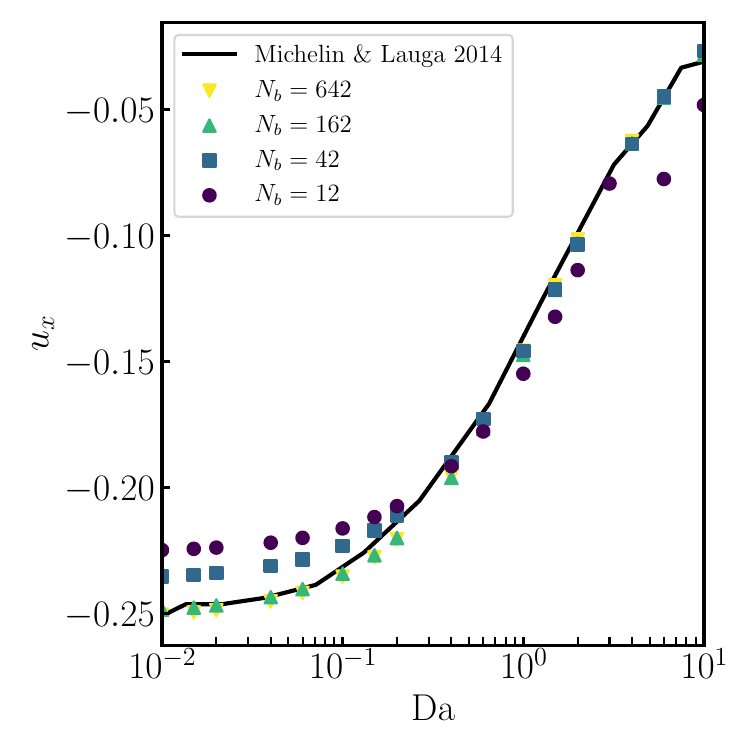}
  \caption{Swimming speed for a Janus sphere versus the Damköhler number.
    Results for resolutions with different number of nodes, $N_b$, and comparison with the results of Michelin and Lauga \cite{MichelinLauga2014}.
    The results with 42 nodes agree well for high Da numbers and shows an error of about $5\%$ for $\text{Da} \ll 1$;
    for finer resolutions the agreement is good for all Da numbers. 
  }
  \label{fig:damkohler}
\end{figure}

We show the swimming speed induced by these surface reactions for several resolutions in Fig.\ \ref{fig:damkohler}.
The general trend is recovered by all our discretizations.
For low Da numbers the concentration cannot diffuse as fast as it is consumed and an asymmetric concentration profile is formed around the sphere.
That asymmetry creates  surface concentration  gradients and consequently a surface slip that induces a swimming speed.
As the Da number increases, and the concentration diffuses faster, the concentration asymmetry is reduced 
which leads to weaker concentration gradients and a swimming speed that tends to zero for $\text{Da} \gg 1$. 
A more careful comparison against the results of Michelin and Lauga shows that our results agree well with a resolution of $N_b=162$ nodes or more.
For 42 nodes we obtain a good agreement for $\text{Da}>1$ and an error of about $5\%$ for $\text{Da} \ll 1$.

For a discretization with only 12 nodes the agreement is worse for all Da numbers.  
Moreover, with that resolution  the velocity exhibits a discontinuity near $Da=4.05$ (not shown),
where the swimming velocity even changes sign.
We postulate that such strange behavior emanates from the nonlinear dependence on the reaction rate, $k$, in the governing equation for 
 the concentration gradient \eqref{eq:grad_c},
which we show more explicitly in  \ref{app:grid_opt_chem}.
Thus, for finite reaction rates we need a finer resolution to resolve the concentration gradient on the colloidal surface with enough accuracy.

\subsubsection{Two emitting particles}
\label{sec:two_emitting}

\begin{figure}[!htb]
  \centering
  \includegraphics[width=0.95\textwidth]{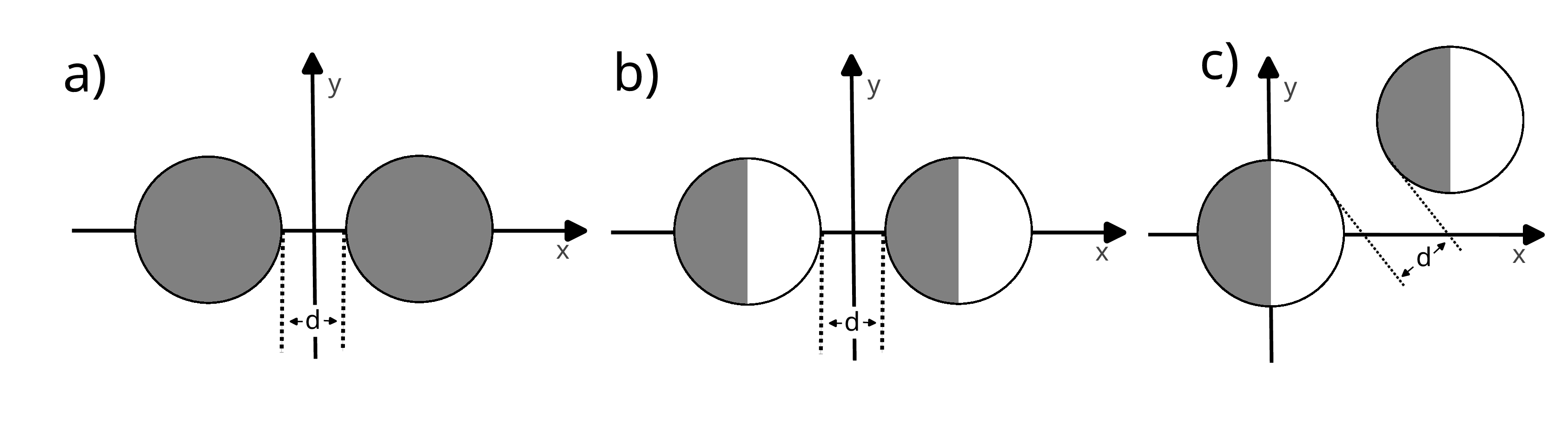}
  \caption{Considered cases to validate the interaction between two active colloids in an unbounded domain.
    {\bf a)} Uniform colloids with $\alpha=1$ and $\mu=1$ on their surfaces.
    {\bf b)} Janus particles with constant mobility $\mu=1$ but active and passive hemispheres, with $\alpha=1$ and $\alpha=0$ respectively,
    shown in grey and white, aligned along their symmetry axes.
    {\bf c)} Janus particles with active and passive hemispheres ($\alpha=\mu=1$ and $\alpha=\mu=0$ respectively) in a non-axisymmetric configuration,
    with the colloids on the line $x=y,\, z=0$.    
    In the three cases the colloids do not consume solute, $k=0$, and are immersed in a constant background field $c_{\infty}=0$.
  }
  \label{fig:two_particles_sketch}
\end{figure}

Here we study the hydrochemical interactions between two active colloids to evaluate the accuracy of the method.
We consider two spherical colloids of unit radius, $R=1$, immersed in an unbounded domain with a constant background field $c_{\infty}=0$.
We consider cases with different production fluxes, mobilities and configurations as sketched in Fig.\ \ref{fig:two_particles_sketch}.
In the three cases considered we solve the Laplace-Stokes problem with optimized discretizations with $N_b=12,\,42,\text{ and } 162$ nodes per colloid,
and we vary the inter colloidal distance.
We compare our results with those obtained by the Diffusiophoretic Force Coupling Method (DFCM), which relies on truncated multipolar expansions, and 
the exact solution computed either in bispherical coordinates for axisymmetric cases or with a highly accurate BEM code for non-axisymmetric configurations \cite{Rojas-Perez2021}.

\begin{figure}[!htb]
  \centering
  \includegraphics[width=0.99\textwidth]{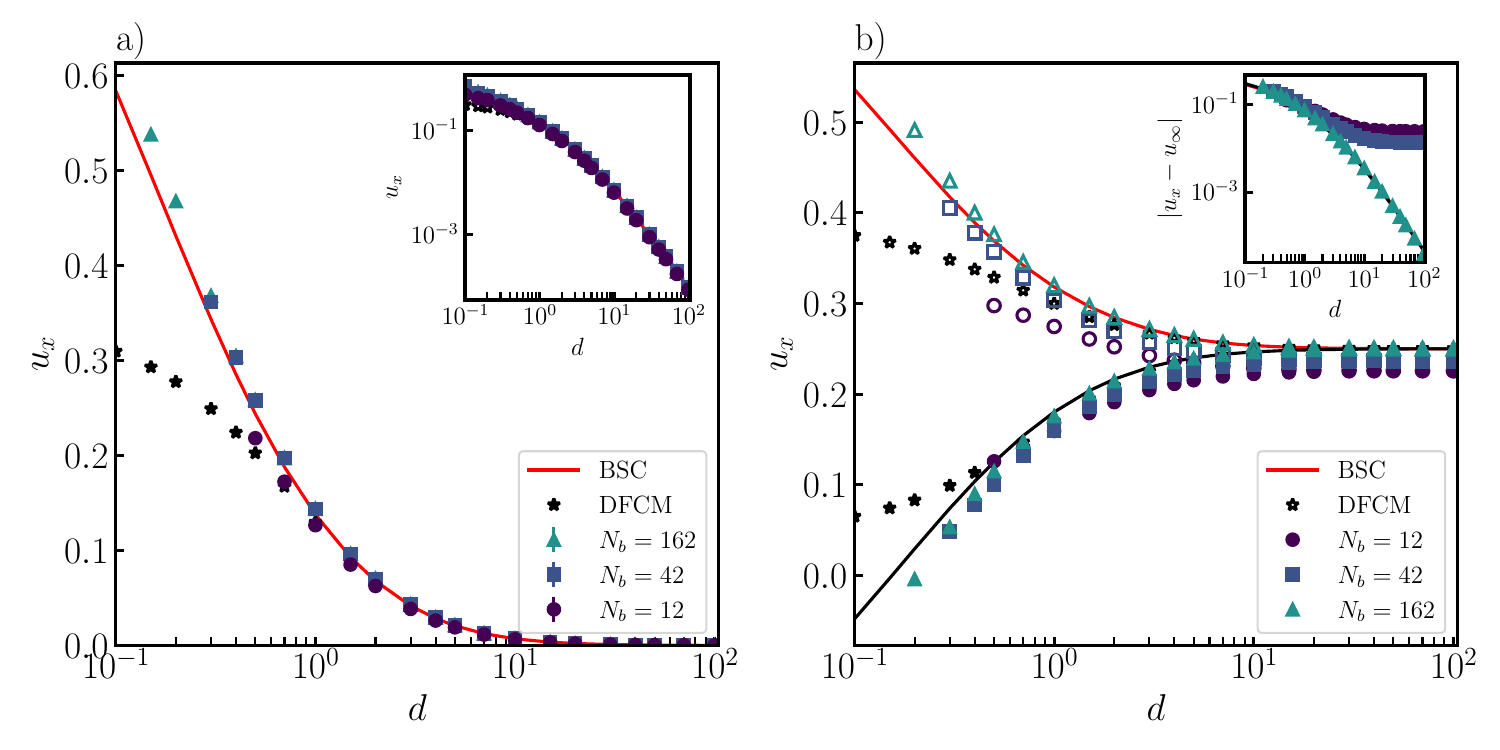}
  \caption{
    Velocities along the $x$-axis for two uniform {\bf (a)} or janus {\bf (b)} colloids as shown in Fig.\ \ref{fig:two_particles_sketch}a and b.
    The motion of the uniform colloids is symmetric thus, we only show the velocity of one of them in {\bf (a)}.
    The motion of the janus colloids is not symmetric and we represent with full and empty symbols the velocity of the leftmost and the rightmost colloid respectively in {\bf (b)}.
    The results for discretizations with $N_b=12,\,42 \text{ and } 162$ are compared with a DFCM and the exact solution BSC.
    We compute the solution down to distances where the nodes of the different colloids start to overlap.
    The low resolution results show an accuracy similar to the DFCM while higher resolutions are more accurate at shorter distances. 
  }
  \label{fig:two_particles}
\end{figure}

In the first setup we consider colloids with a uniform emission flux and mobility, $\alpha=1$ and $\mu=1$, on their surfaces.
Since the discretized colloids are not perfectly isotropic, see Fig.\ \ref{fig:squirmer_speed_optimized}a, we only compute the solution down to distances where the nodes, of radius $a$, of the different colloids start to overlap.
This way the results do not depend on the relative orientation of the colloids.
Since the node radius decreases for finer grids, higher resolutions overlap at a smaller gap size.
We show in Fig.\ \ref{fig:two_particles}a the velocity for one of the colloids as a function of the gap $d$ between the two colloids.
Close colloids generate a concentration gradient in their gap which induces a phoretic speed.
As the particles are moved apart the magnitude of the concentration gradient decreases and the speed decays to zero.
We observe that our lowest resolution, $N_b=12$, is accurate down to gaps of one colloidal radius, $d=1$, just as the DFCM.
For higher resolutions the solution remains accurate for smaller gaps and it surpasses the accuracy of the DFCM.

In our second test we use two Janus colloids, with emitting fluxes $\alpha=1$ and $\alpha=0$ on each hemisphere, and a uniform mobility $\mu=1$,  forming an axisymmetric configuration
as shown in Fig.\ \ref{fig:two_particles_sketch}b. 
The velocity is shown for three discretizations in Fig.\ \ref{fig:two_particles}b.
A single Janus colloid is able to swim due to the self-generated concentration gradient.
Thus, unlike the previous case, this problem is not symmetric.
Fig.\ \ref{fig:two_particles}b shows that for close colloids, i.e.\ 
gaps below 1 colloidal radius, our discretizations with 42 or 162 nodes are more accurate than the DFCM, while the 12 nodes discretization is somewhat less accurate.
For very large distances the colloids attain the swimming speed of an isolated colloid, $\bu^{\infty} = \be_x / 4$.
The DFCM is specially designed for spherical particles and is tuned to recover this result exactly, thus its error decays to zero when the colloids move apart.
Our discretization works for arbitrary colloidal shapes at the cost of introducing a discretization error
that affects even isolated colloids.
Thus, the error does not decay to zero at long distance as it can be seen in the inset of Fig.\ \ref{fig:two_particles}b.
These errors are around $9.7\%$, $5.4\%$ and $0.02\%$ for the discretizations with 12, 42 and 162 nodes respectively.
The error is small for moderate resolutions and the solution can be refined.

\begin{figure}[!htb]
  \centering
  \includegraphics[width=0.99\textwidth]{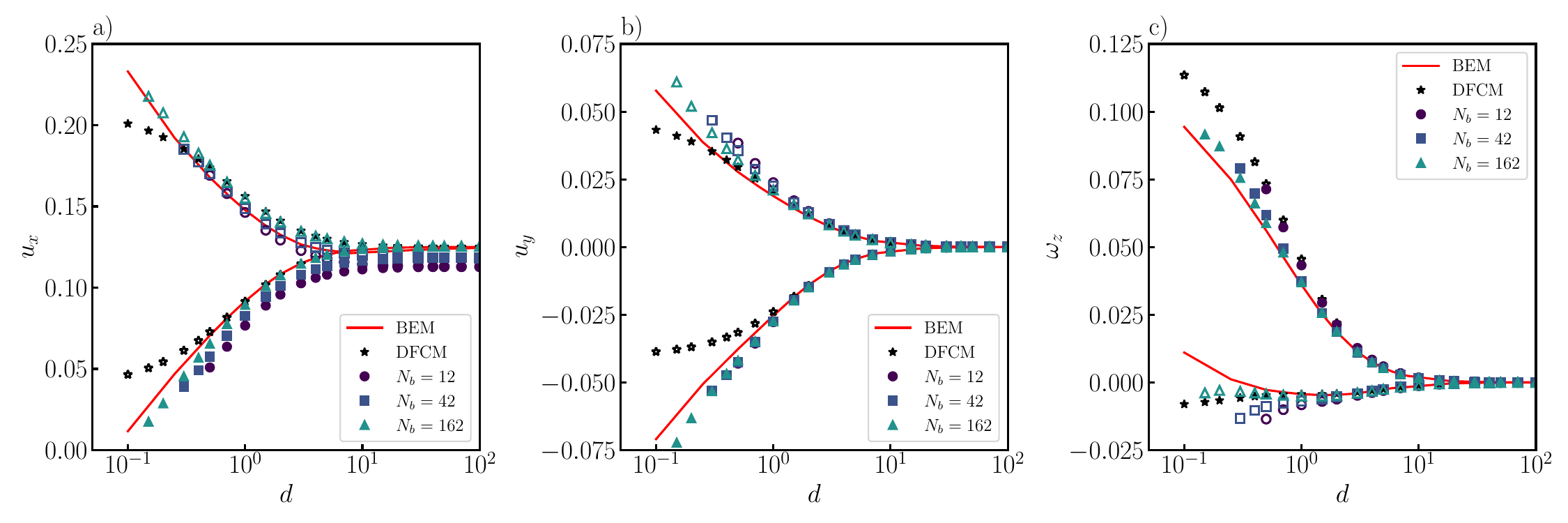}
  \caption{
    Numerical results for a pair of Janus colloids in an non-symmetric configuration as shown in Fig.\ \ref{fig:two_particles_sketch}c.
    We represent with full and empty symbols the velocity of the rightmost and leftmost colloid.
    The results for discretizations with $N_b=12,\,42 \text{ and } 162$ are compared with a DFCM and essentially the exact solution computed with a BEM.
    The three panels, {\bf a, b,} and {\bf c}, show the linear velocity along the $x$ and $y$ axes and the out of the plane angular velocity respectively.
  }
  \label{fig:two_janus_non_symmetric}
\end{figure}

As a final test we consider two Janus colloids where the active and passive hemispheres have different reaction fluxes and surface mobilities,
$\alpha=\mu=1$ and $\alpha=\mu=0$, in an asymmetric configuration as shown in Fig.\ \ref{fig:two_particles_sketch}c.
By symmetry, the colloids remain in the $z=0$ plane but acquire velocities in the $x$ and $y$ directions as well as an angular velocity directed along the $z$ axis.
Our results, compared with   the DFCM method and a highly accurate BEM solution, are shown in Fig.\ \ref{fig:two_janus_non_symmetric}.
In general our results are more accurate than the DFCM for short distances with all the discretizations considered.
The $y$ component of the velocity and the angular velocity decay to zero for large distances as expected.
As before, the $x$ component of the velocity shows the same finite error for large intercolloidal distances, but it can be controlled by increasing the resolution.
Overall, these tests show that using a discretization with $42$ nodes the numerical results remain reasonably  accurate down to
interparticle distances of $0.3$ colloidal radius.

\section{Simulations}
\label{sec:Simulations}

In this section we use the flexibility and accuracy of the method to investigate more complex systems with large numbers of particles with intricate shapes. The first example focuses on the individual and collective motion of gravitactic active rods down an incline. The second system consists of chiral particles that rotate and repel each other due to hydrochemical interactions. Both examples explore new physical problems that, to the best of our knowledge, have not been addressed with numerical simulations before. 
\subsection{Gravitactic active rods}
\label{sec:Simulations_rods}
Thanks to their ease of manufacture, rod-like phoretic particle have been widely investigated experimentally over the last two decades. 
Inspired by the recent work of  Brosseau \textit{et al.} \cite{Brosseau2021}, we investigate the dynamics of colloidal tail-heavy Janus rods with asymmetric surface properties swimming on an incline, with a tilt angle $\alpha = 20^{\circ}$. In their work they have shown that isolated swimmers with a heavier tail were gravitactic and thus oriented against gravity. In addition they tilted head-down due to hydrochemical interactions with the wall underneath, which induced a fore-aft drag asymmetry and facilitated their upward swimming along steep slopes.

In this section, we focus on the collective dynamics of such swimmers in order to evaluate the effect of multi-body hydrochemical interactions on their ability to swim uphill. Our rods, with length $L = 4 \mu$m and radius $R = 1 \mu$m,  are discretized with $N_b = 110$ nodes on their surface as described in Section \ref{sec:smooth-cyl-valid}. They are immersed in a fluid with viscosity similar to water $\eta = 1 $mPa$\cdot$s at temperature $T = 300$K ($=26.9^{\circ}$), with a constant background fuel concentration field $c_{\infty} = 100 \si{\mu m^{-3}}$. They 
 consume fuel asymmetrically at a rate $k_B = 2 \mu$m$\cdot$s$^{-1}$ at the back and $k_F = 0 \mu$m$\cdot$s$^{-1}$ at the front. The mobility is also asymmetric with $\mu_B = -4 \mu$m$^5\cdot$s$^{-1}$ and $\mu_F = 0 \mu$m$^5\cdot$s$^{-1}$, and the solute diffusivity is $D = 1 \mu$m$^2\cdot$s$^{-1}$.
The resulting Damkholer number is Da $= \bar{k}L/D = (k_F + k_B)L/2D = 4$. 
In our simulations we exaggerate the role of gravitaxis and consider a tail eight times heavier than the tip, where the total excess  mass (compared to the solvent) of the rod is  $m_{\text{e}} = 6.8\times 10^{-9}$mg. 
In the short range, the rods repel each other and/or the wall below with a repulsive potential to prevent overlaps.

\begin{figure}[!htb]
  \centering
\includegraphics[width=\textwidth]{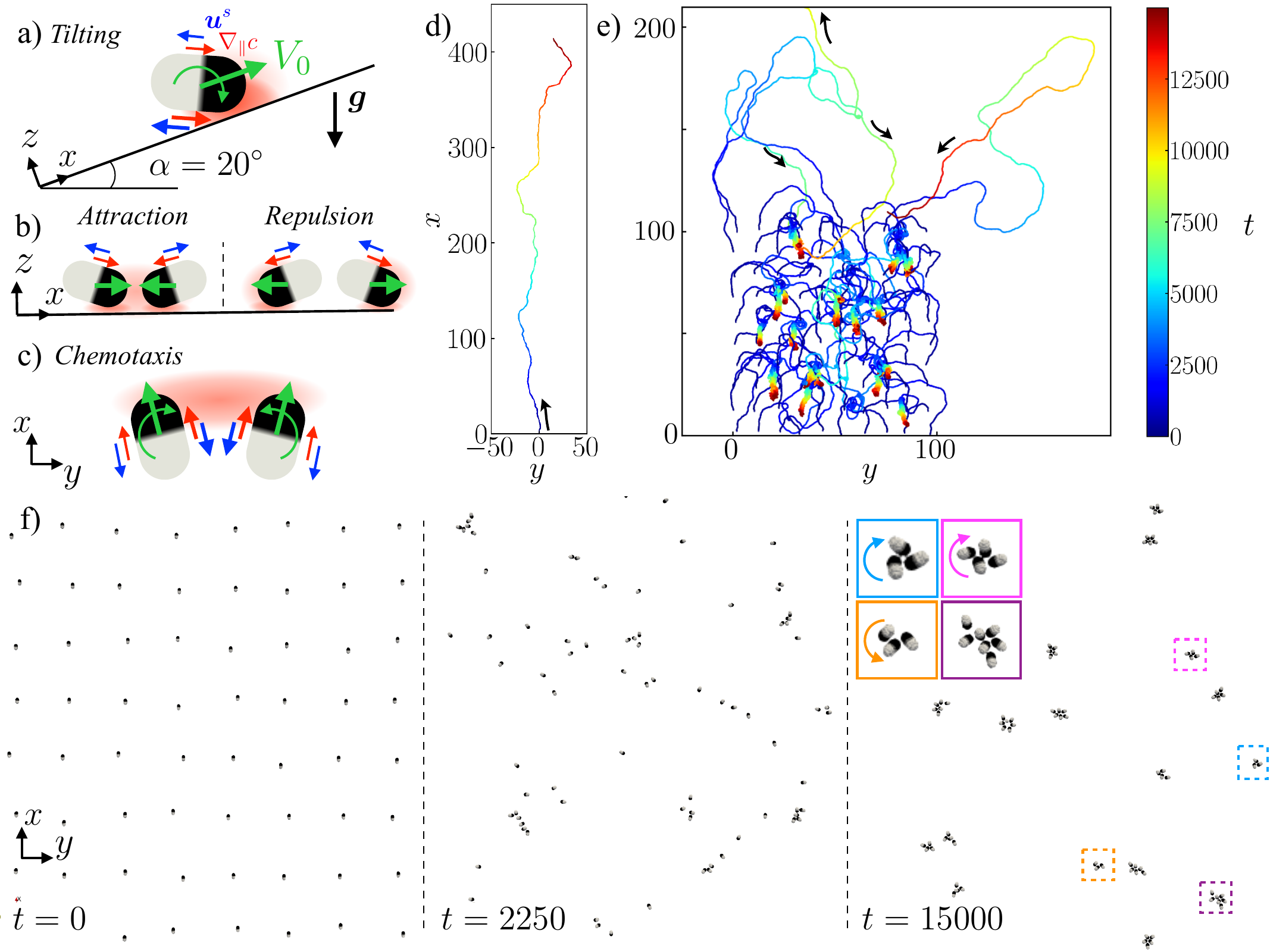}
\caption{ a) Sketch of the self-propulsion and tilting mechanisms of a phoretic rod above an incline with angle $\alpha = 20^{\circ}$.
  Red and blue arrows represent concentration gradients and slip velocities respectively.
  b) Basic mechanism of phoretic attraction between the inert parts  and repulsion between the reactive parts. c) Mechanism of chemotactic reorientation. d) Trajectory of an isolated rod swimming uphill on an incline (top view). e) Trajectories of 100 rods initially placed on a square lattice. Colorbar: dimensionless time. f) Snapshots of the simulation over time. Insets in the last panel: close-up view on some clusters, colored arrows show their rotation direction if any.  See also Supplemental Movie 1. }
  \label{fig:100_smooth_cyl}
\end{figure}

First, we investigate the motion of a single Janus rod initially oriented along the $x$-axis.
In the absence of walls, the concentration is higher near the inert part, where no fuel is consumed. The resulting concentration  gradient is directed from the inert to the reactive side, which, by phoresis, generates a surface slip flow in the opposite direction, and therefore, by momentum conservation, a straight swimming motion along the $x$-axis with the inert part first. 
In the presence of a wall, as sketched in Fig.\ \ref{fig:100_smooth_cyl}a, the concentration field is highest below the tip of the inert part due to confinement. The resulting concentration gradient, and thus the slip flow, is therefore stronger below  than above the rod. This, together with a smaller relative pressure ahead of the slip region \cite{Brosseau2021}, leads to a reorientation of the rod with its inert part toward the wall, with a net motion uphill along the $x$-axis. In the absence of thermal fluctuations, the Janus rod swims uphill at a steady speed $V_0 = 6.26 \mu$m$\cdot$s$^{-1}$. However, due to their small size, $O(\mu$m$)$, these particles are sensitive to the Brownian motion of the solvent molecules. As shown in Fig.\ \ref{fig:100_smooth_cyl}d, a typical rod trajectory fluctuates significantly but the rod  swims uphill in average due to the gravitactic torque that reorients it against gravity. Similar trajectories have been observed experimentally with bimetallic rods \cite{Brosseau2021}.

Now we consider a large collection ($M = 100$) of such rods initially placed on a lattice at a distance $d_x = d_y = (10 \pm 2.5)L$ from each other and oriented  along the $x$-axis. The equations of motion are integrated with the Euler-Maruyama traction scheme 
 which requires solving the Stokes linear system \eqref{eq:linear} twice per time-step \cite{Sprinkle2017}.  We simulate the suspension for 30000 time-steps, with a time-step size $\Delta t = 0.016 L/V_0$, which corresponds to the time for an isolated rod to travel $\approx 470$ its body length $L$, and it took 3.5 days on a 10-core computer.
Fig.\ \ref{fig:100_smooth_cyl}e-f shows the trajectory of the rods' center of mass and three corresponding snapshots, where lengths are rescaled with $L$, speed with $U_0=\bar{\mu}\bar{k}/D$ and time with $L/U_0$ (see also Supplemental Movie 1). \\
Unlike the isolated swimmer in panel d, all particles, except for one that manages to escape, reorient toward each other and eventually merge into small clusters that slowly sediment down the incline. Some clusters spontaneously rotate, and change their rotation direction due to geometrical rearrangements induced by thermal fluctuations, while others barely move because of their symmetry. A closer look at the clusters show that the particles point inwards with their inert part. This is because the inert parts are mutually phoretically attractive, while the reactive parts are phoretically repulsive. Indeed, as sketched in Fig.\ \ref{fig:100_smooth_cyl}b, the concentration gradient, and thus particle motion, is always directed towards the inert parts. 
These clusters form because particles reorient and swim toward each other. This reorientation results from the competition between gravitaxis, that tends to orient particles against gravity, i.e.\ with their inert part directed along the $x$-axis, and chemotaxis that tends to orient particles toward chemical gradients. The effect of chemotaxis is sketched in Fig.\ \ref{fig:100_smooth_cyl}c: due to confinement, there is a concentration build up between the inert parts of two neighbouring particles, this local increase generates a stronger gradient in the space between the particle pair, and thus rotates the inert parts toward each other. However, once formed, these clusters do not grow indefinitely. We postulate it might be due to the phoretic repulsion between the reactive parts of the rods that are pointing outwards from the clusters. A more detailed study of the clustering dynamics of anisotropic phoretic particles will be the subject of future work.

 \subsection{Collective motion of chiral phoretic  microrotors}
 \label{sec:Simulations_microrotors}
\begin{figure}[!htb]
  \centering
\includegraphics[width=\textwidth]{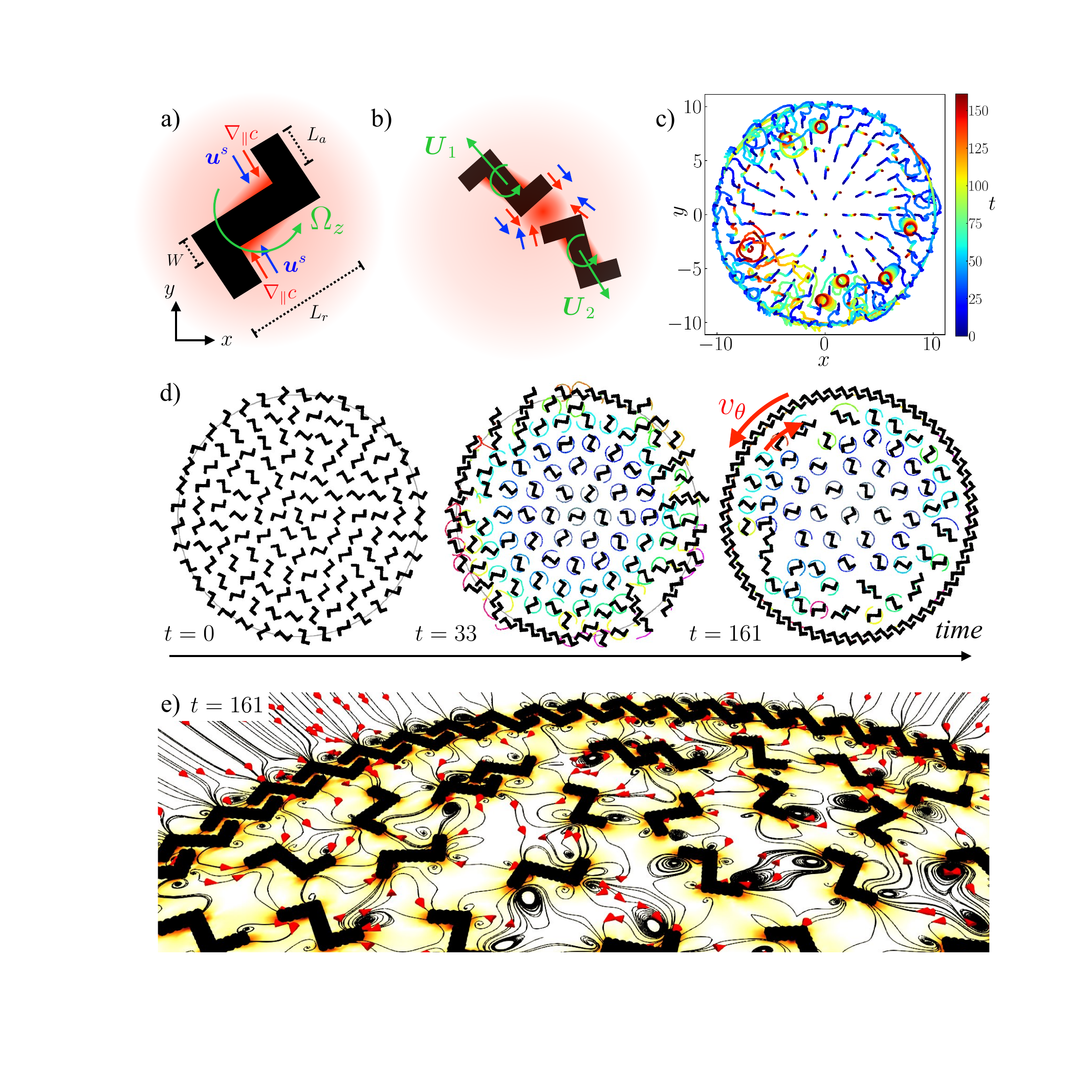}
  \caption{a) Geometry of a chiral particle and schematic of the self-rotation mechanism.  b) Phoretic repulsion between two chiral particles with positive mobilities ($\mu>0$). c) Trajectories of the particles center of mass over time (colorbar). d) Snapshots of the simulation at three different times. Light grey circle: confining disk of radius $R = 10$. Colored lines: trajectory of the tip of the arms over the last rotation period. Red arrows: azimuthal motion, $v_{\theta}$, of the particles at the periphery. See also Supplemental Movie 2. e) Flow field around the chiral particles at $t=161$. Color-scale from white to dark red: magnitude of the velocity field. Black lines with red arrows: streamlines. }
  \label{fig:130_sprinklers}
\end{figure}

In order to design functionalized  colloidal systems and materials, one must control the self-induced motion of different active particles in a common environment. Recent efforts by Brooks \textit{et al.} \cite{Brooks2019}, Zhang \textit{et al.} \cite{Zhang2019} and Sharan \textit{et al.} \cite{Sharan2021} have shown that, just like translation, the rotation of catalytic particles can be controlled with their shape. 
 In their work, they control both the magnitude and direction of rotation of an individual particle, by breaking its rotational symmetry with its chirality. While the phoretic mechanism leading to spontaneous rotation of a single particle is quite intuitive and well understood, their interactions and collective motion at large scale have not yet been investigated.\\
 Here we use our new framework to simulate, for the first time to our knowledge, the motion of 130 chiral phoretic particles circularly confined by a harmonic potential above a wall. 
 Our chiral particles simply consist of one main rod of length $L_r = 9 \mu$m and width $W = 1 \mu$m, with two arms of length $L_a = 4 \mu$m and width $W$  placed asymmetrically at its extremities, as shown in Fig.\ \ref{fig:130_sprinklers}a. Each  particle has a uniform surface mobility $\mu = 1 \mu$m$^5 \cdot$s$^{-1}$  and emits solute uniformly at a fixed flux $\alpha  = 1 \mu$m$^{-2}\cdot$s$^{-1}$. The solute diffusivity is chosen arbitrarily as $D = 1 \mu$m$^2\cdot$s$^{-1}$. 
  The particles are discretized with a low resolution grid made of $N_b = 72$ adjacent nodes of radius $a = W/2 =  0.5 \mu$m. This minimal model might not describe accurately the flow and concentration field at  the surface of the body but still captures the main mechanisms described below: self-induced rotation and phoretic repulsion.

 The mechanism for self-induced rotation is schematized in Fig.\ \ref{fig:130_sprinklers}a.
 The particle releases solute uniformly over its surface, but due to its bent shape, the solute  primarily accumulates near the inner corners of the arms. This accumulation generates a sharp gradient along the arms, which, for a positive mobility coefficients $\mu>0$, induces a slip flow $\bu_s$ towards the body. To preserve momentum the arms move against this phoretic flow, which, by symmetry,  leads to the rotation of the whole body about the $z$-axis at a constant rate $\Omega_z$. 
When another chiral particle is nearby, see Fig.\ \ref{fig:130_sprinklers}b, the solute also accumulates between the particles' boundaries. The resulting solute gradients induce phoretic flows which, for $\mu>0$, push the particles away from each each other while rotating. This phoretic repulsion mechanism also repels the particles away from the floor underneath.

 To investigate their collective motion we initially place 130 particles in a disk of radius $R = 10 L_r$  as shown on the first panel of Fig.\ \ref{fig:130_sprinklers}d.  The particles are denser than the fluid and sediment near the floor at an equilibrium height $h = 0.24L_r$ that results from the balance between gravity, phoretic repulsion and electrostatic repulsion from the floor.  The particles are confined in the disk by a harmonic potential of the form 
 $$U_{\text{disk}}(r) = \fr{k_U}{2}\pare{r-R}^2, \, \text{if } r>R$$
 where $r$ is the radial distance between a particle and the center of the disk and $k_U$ the stiffness of the potential. 
 Steric interaction between rigid bodies and electrostratic repulsion from the wall are accounted for with a short-range pairwise repulsive potential between the nodes and between the nodes and the wall.
 In the following, lengths are nondimensionalized with the particle length $l_c = L_r$, velocities with $U_c  = \alpha \mu/D$ and time with $t_c = l_c/U_c$. Here we neglect the effect of thermal fluctuations. 
 The equations of motion are time-integrated with a simple explicit Euler scheme and the time step is  $\Delta t \approx 2 \cdot 10^{-4} T_{\text{rot}}$, where $T_{\text{rot}} = 2\pi/\Omega_z$ is the time for a full rotation of an isolated particle. We simulate the system for $145 000$ time-steps, which corresponds to $\approx 30$ body rotations, and it took 5.5 days on a 10-core computer. 
 
Figure \ref{fig:130_sprinklers}c shows the trajectory of the particles center of mass over time
%, where a collective radial migration is visible at short time and a slow azimuthal  drift happens at longer time. Some trajectories also exhibit circular shapes, which result from the formation of clusters rotating as rigid bodies. 
and Figure \ref{fig:130_sprinklers}d shows the corresponding snapshots at three different times (see also Supplemental Movie 2). 
A collective radial migration happens at short time due to phoretic repulsion between the particles: the particles repel each other and migrate towards the outer edge of the disk while rotating. Due to the circular confinement of the harmonic potential, they accumulate on the rim and  progressively form of a tightly packed circular ring with no defects.
As it self-assembles, the ring rotates counter-clockwise (CCW). 
The particles inside the disk progressively stabilize their radial position. Some of them form intermittent chains, which rotate like rigid bodies, and gradually break into monomers and dimers. Due to the competition between phoretic repulsion and harmonic confinement,  the system seems to converge to a crystalline structure in the bulk which coexists with a few persistent dimers. At steady state, the inner particles barely move, while the ones near the ring  slowly migrate in clockwise (CW) direction (see trajectories in Fig.\ \ref{fig:130_sprinklers}c and red arrows on panel d).

\begin{figure}[!htb]
  \centering
\includegraphics[width=0.85\textwidth]{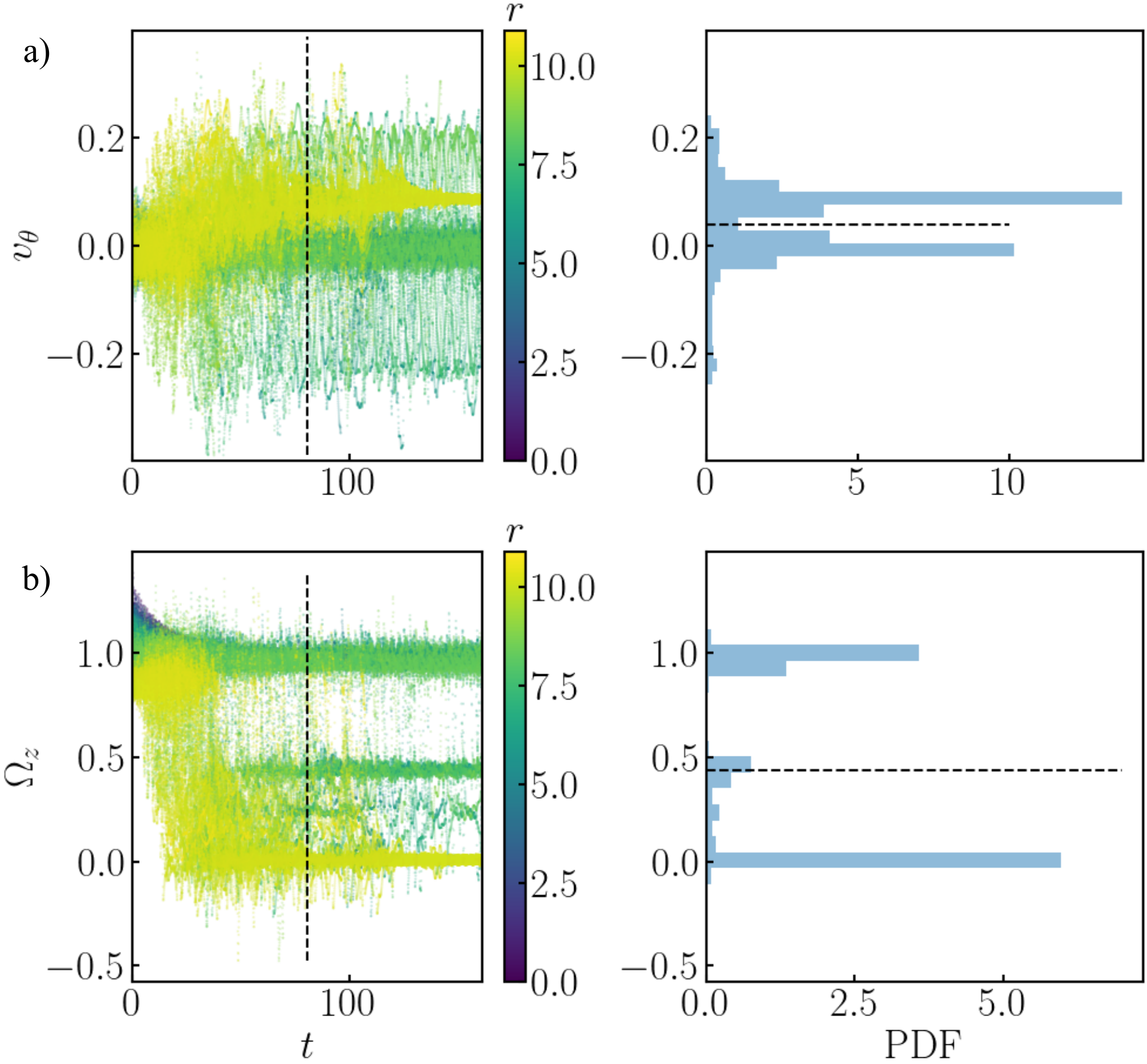}
  \caption{Time evolution (left) and probability density distribution (right) of the azimuthal velocity $v_{\theta}$  (a) and rotation rate $\Omega_z$ (b) of the chiral particles. Colorbar: radial position of the particles. Dashed line: onset of steady state (left), and average value of the PDF (right). See also Supplemental Movie 3.}
  \label{fig:130_sprinklers_stats}
\end{figure}

The azimuthal motion in the system is reported in Fig.\ \ref{fig:130_sprinklers_stats}a. The time evolution and the probability density function (PDF) of  the azimuthal velocity $v_{\theta}$ show that the CCW motion of the particles on the ring stabilizes around $v_{\theta}\approx 0.08$ while the particles inside mostly move CW ($v_{\theta}<0$) but very slowly $|v_{\theta}|\ll 0.08$. Even though the distribution is bimodal, the mean azimuthal motion in the system is CCW (see dashed line at  $\bar{v}_{\theta} \approx 0.04$).

Rotational motion is also very different between the ring and bulk particles. 
 As shown in the left panel of Fig.\ \ref{fig:130_sprinklers_stats}b and on Supplemental Movie 3, after a transient time, $t>80$, geometrical frustration prevents the particles on the ring from rotating ($\Omega_z = 0$ at $r=R = 10$), while the ones at the center keep a steady rotation rate $\Omega_z \approx 1$, except for the few particles forming chains/dimers at the periphery ($\Omega_z \leq 0.5$). As shown by the PDF on the right panel, these chains are scarce compared to the two large peaks at $\Omega_z = 1$ and $\Omega_z = 0$ \\

\section{Discussions}
\label{sec:discussions}

In this work we have presented a numerical method to solve the hydrochemical problem that appears in many soft matter systems 
where colloids  mediate chemical reactions on their surface and interact both chemically and hydrodynamically. 
The strengths of our approach are its flexibility, it allows to simulate complex shaped colloids in different domains, 
the possibility to include Brownian motion, 
the easiness of implementation and a moderate computational cost. 
These advantages come at the cost of a moderate accuracy, typically two or three digits accurate. 
Nonetheless, the accuracy is controllable if one is willing to accept higher computational costs. 
The error in the chemical and hydrodynamic interactions between particles can also be reduced using pair-corrections \cite{Broms2023}.

Our method only discretizes the surface of the colloids with simple quadrature rules which helps simulating large systems. 
To solve the Stokes equations we use a regularized mobility, the so-called RPY mobility, which is always positive definite and thus it eases the generation of the Brownian noise
\cite{Usabiaga2016, Sprinkle2017}. 
To solve the Laplace problem we use a simple quadrature where we eliminate the self-interaction of the nodes.
We use the same grid to solve both problems so the surface slip on phoretic particles can be easily computed and plugged in the Stokes problem.

All the steps to solve the coupled Laplace-Stokes problem are matrix free and allow the use of fast methods (e.g.\ the Fast Multipole Method \cite{Yan2020})
to compute the hydrochemical interactions. 
Moreover, the linear systems to solve both problems converge in a moderate number of iterations that scales weakly with the number of colloids $M$ for the Laplace problem  ($\mc{O}\pare{M^{1/5}}$), and is independent of $M$ for the Stokes problem \cite{Usabiaga2016}.

In this work, we have optimized the grid only for the hydrodynamic part of the problem. 
However, the optimal grid for the Stokes problem might not always be optimal for the Laplace problem. 
In future work we plan to optimize the grid for both problems simultaneously or to use a different grid for each problem. 
In \ref{app:grid_opt_chem} we provide some ideas and directions that will be tested to match the linear operators of the Laplace problem.

With this approach we have been able to simulate 100 phoretic rods on an inclined plane in Sec.\ \ref{sec:Simulations_rods}
or 130 chiral colloids above a floor in Sec.\ \ref{sec:Simulations_microrotors}. 
In addition to their visual appeal (Fig.\ \ref{fig:130_sprinklers}e), these first results open many interesting questions and perspectives regarding the design and collective motion of chemically powered  colloidal machines. Besides  capturing the main hydrochemical mechanisms at the individual level, our tool  predicts their complex dynamics at large scales. 
In the future it would be interesting to mix particles with different surface properties or particles with opposite chiralities (i.e.\ opposite rotation directions) to investigate their ability to mix or  phase-separate, like binary mixtures of torque-driven particles \cite{Yeo2015}.  
Finally, the effect of the Damkholer number on propulsion has only been investigated for isolated particles so far. Since our approach is valid for any value of Da, it would also be interesting to the study the effect of Da on the collective dynamics of reactive suspensions.

Our code is open source and freely available on GitHub (\url{https://github.com/stochasticHydroTools/RigidMultiblobsWall}). The online repository also includes the grid optimization routine described in Section \ref{sec:Grid_opt} and some documented examples shown in the paper: a single active sphere (Section \ref{sec:damkholer}) and the chiral microrotors (Section \ref{sec:Simulations_microrotors}). 
To conclude, we would like to emphasize that our framework is not limited to diffusiophoresis, i.e.\ solute concentration fields. It broadly applies to any scalar field that satisfies elliptic equations, such as temperature or electric potential. With minor modifications, our code can be used to study the dynamics of thermophoretic or electrophoretic particle suspensions.

\section*{Acknowledgments}
We warmly thank Anna Broms for sharing the code to discretize the smooth rods shown in Section \ref{sec:smooth-cyl-valid}.
F.B.U.\ acknowledges support from ``la Caixa'' Foundation (ID 100010434), fellowship LCF/BQ/PI20/11760014, and from the European Union's Horizon 2020 research and innovation programme under the Marie Skłodowska-Curie grant agreement No 847648
and also by the Basque Government through the BERC 2022-2025 program and by the Ministry of Science, Innovation and Universities:
BCAM Severo Ochoa accreditation CEX2021-001142-S/MICIN/AEI/10.13039/501100011033.
B.D.\ acknowledges support from the French National Research Agency (ANR), under award ANR-20-CE30-0006.  B.D.\ also thanks the NVIDIA Academic Partnership program
for providing GPU hardware for performing some of the simulations reported here.

\appendix

\section{Scaling of the singular values of the slip mobility matrix}
\label{app:scalingSV}

\begin{figure}[!htb]
  \centering
  \includegraphics[width=0.49\textwidth]{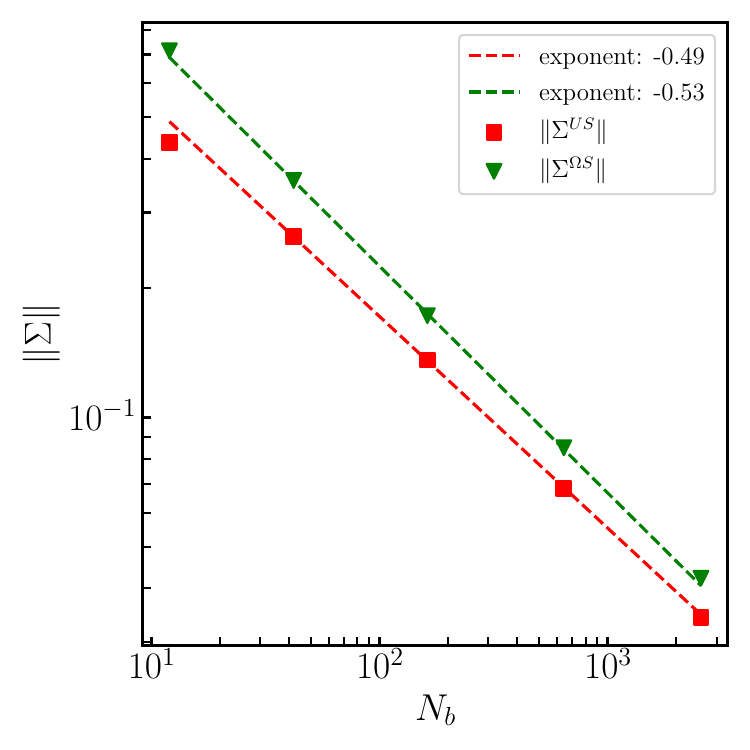}
  \caption{Scaling of the singular values of each block of the slip mobility matrix with the number of nodes $N_b$.}
  \label{fig:scaling_sigma_nblobs}
\end{figure}

As shown in Fig.\ \ref{fig:scaling_sigma_nblobs}, the singular values of each block scale as $N_b^{-1/2}$ with the rigidmultiblob method (and presumably with any numerical method). 
To understand this scaling, first note that the particle velocity resulting from the slip, $\bU = -\wtil{\bN}\bu_s$  does not scale with the number of surface nodes $N_b$. The $6 \times 6$ matrix $\bW$ in the SVD of $\wtil{\bN}$ contains  $r = 6$ basis vectors of the column space of $\wtil{\bN}$, where $r$ is its rank. Since $r$ and the magnitude of the basis vectors do not depend on $N_b$, neither does $\bW$. \\
On the other hand, the matrix $\bV$ contains $3N_b$ column vectors, $\bv_i, i=1,..,3N_b$, with $3N_b$ components each. Each of these column vectors is unitary so that $\bv_i \sim N_b^{-1/2} \breve{\bv}_i$, where $\breve{\bv}_i$ is the same vector without normalization (indeed  $\|\bv_i\| = \left(\sum^{3N_b}_{j=1} v_{i,j}^2\right)^{1/2} = 1$).
As a result, the dot product between one of these vector and $\bu_s$ is $\bv_i\cdot\bu_s  = \sum^{3N_b}_{j=1} v_{i,j}u_{s,j} \sim N_b^{-1/2}   \underset{\sim N_b}{\underbrace{\sum^{3N_b}_{j=1}\breve{v}_{i,j}u_{s,j}}} \sim N_b^{1/2}$, so that 
\eqn{
\bV^T\bu_s \sim \left[\begin{array}{c}
N_b^{1/2}\\
\vdots\\
N_b^{1/2}\\
\end{array}\right] = N_b^{1/2} \left[\begin{array}{c}
O(1)\\
\vdots\\
O(1)\\
\end{array}\right].
}
Thus, the velocity $\bU = -\wtil{\bN}\bu_s = -\bW\bSigma\bV^T\bu_s$ can be written as
\eqn{
\bU \sim - \bW\cdot\left[\begin{array}{c}
\sigma_1 N_b^{1/2}\\
\vdots\\
\sigma_6 N_b^{1/2}\\
\end{array}\right].
}
Since $\bU$ and $\bW$ do not scale with $N_b$ then the singular values must scale as $\sigma_i \sim N_b^{-1/2},\, i=1,..6$.

\section{Optimized grid with a different cost function}
\label{app:minmax}

\begin{figure}[!htb]
  \centering
  \includegraphics[width=0.49\textwidth]{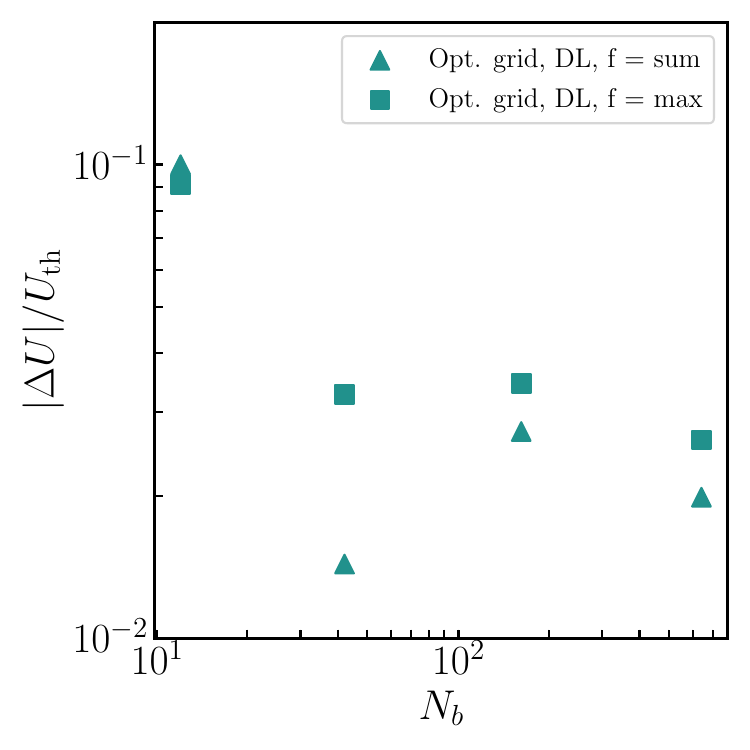}
  \caption{Relative error of the propulsion speed of a squirmer as a function of the grid resolution $N_b$ for two optimal grids obtained with different cost functions: \eqref{eq:cost_func} using the sum of the errors (triangles),  \eqref{eq:cost_func_max} using the $\max$ of the errors (squares).}
  \label{fig:squirmer_speed_optimized_cost_funcs}
\end{figure}

In this Appendix we use another cost function for the grid optimization problem \eqref{eq:opt_pb}:
\eqn{
f(S,a) = \max\{E_{UF}, E_{\Omega F}, E_{UT}, E_{\Omega T}, E_{US}, E_{\Omega S}\}
\label{eq:cost_func_max}
}
which is the one used by \cite{Broms2023} to match the mobility coefficients of an ideal sphere. We have found that in the absence of slip, both cost functions \eqref{eq:cost_func} and \eqref{eq:cost_func_max} provide the same optimal grid parameters. However, as shown in Table \ref{tab:opt_res_minmax}, when adding the slip mobility matrix, the two cost functions lead to slightly different optimal grids, which, in turn, results in different swimming speeds, where, as shown in Figure \ref{fig:squirmer_speed_optimized_cost_funcs}, the optimal grid using the sum of the errors \eqref{eq:cost_func} performs better than the one taking the $\max$ error \eqref{eq:cost_func_max}.

\begin{table}[htb!]
\begin{center}
\begin{tabular}{ c | c c | c c } 
&  \multicolumn{2}{c|}{Opt.  grid DL \eqref{eq:cost_func}} &  \multicolumn{2}{c}{Opt.  grid DL \eqref{eq:cost_func_max}} \\
 \hline
 $N_b$ & $S$  & $a/s$ & $S$  & $a/s$ \\
 \hline
12&  0.858 & 0.412  &  0.831 & 0.406\\
42&  0.929 &  0.410 & 0.913 &  0.433\\
162&  0.965 & 0.413 & 0.959 & 0.429\\
642&   0.982 & 0.424&  0.980 & 0.480\\
\end{tabular}
\caption{ Optimal grid parameters  obtained with two different cost functions \eqref{eq:cost_func} and \eqref{eq:cost_func_max} respectively  using the DL formulation for a sphere with radius $R = 1$ in an unbounded domain.}
\label{tab:opt_res_minmax}
\end{center}
\end{table}

% \section{Complete formulation}
\section{Half space Green's functions}
\label{sec:green}
The  Laplace Green's function and dipole kernel in a half space bounded by a impervious solid wall at $z=0$ are
\eqn{
  \label{eq:kernel_monopole}
  G(\bx,\by) &= \fr{1}{4\pi} \fr{1}{r} + \fr{1}{4\pi} \fr{1}{r^*} , \\
  T_i(\br) &= -\fr{1}{4\pi}\fr{r_i}{r^3} - \pare{1 - 2 \delta_{i3}}  \fr{1}{4\pi}\fr{r^*_i}{(r^*)^3}, 
}
with $\br=\bx-\by$ and $\br^*=\bx - \bP \by$ with $\bP=\bI - 2\be_z\be_z^T$.
Thus, the Green's function \eqref{eq:kernel_monopole} has an additional source at the other side of the wall.
When we take the derivative of \eqref{eq:complete_v1} to find the gradient, we derive with respect the target position $\bx$.
Therefore, in the operators of \eqref{eq:grad_c} we have to use the kernels
\eqn{
  T_i(\br) &= -\fr{1}{4\pi}\fr{r_i}{r^3} - \fr{1}{4\pi}\fr{r^*_i}{(r^*)^3}, \\
  L_{ij}(\br) = \fr{\partial}{\partial x_i} T_j(\bx,\by) &= -\fr{1}{4\pi} \fr{\delta_{ij}}{r^3} + \fr{3}{4\pi}\fr{r_i r_j}{r^5}
  -\pare{1 - 2 \delta_{j3}} \corchete{\fr{1}{4\pi} \fr{\delta_{ij}}{(r^*)^3} - \fr{3}{4\pi}\fr{r^*_i r^*_j}{(r^*)^5}}.  
}
Thus, the operator $L_{ij}$ is not symmetric any more.

\section{Some ideas on grid optimization for the Laplace problem}
\label{app:grid_opt_chem}
In the discrete setting, the concentration and its gradient on the particle surface due surface fluxes $\balpha$, reaction rates $\bk$, background field $\bc_{\infty}$ and background gradient $\bna c_{\infty}$ at the node locations are given by \eqref{eq:complete_discrete}
\eqn{
  \label{eq:complete_v3}
  \underset{\bA}{\underbrace{\corchete{\fr{1}{2} \bI \bw^{-1} + \bD + \bS \fr{\bk}{D}} \bw}} \bc = \bS \fr{\bw \balpha}{D} + \bc_{\infty},
  %\underset{\bA}{\underbrace{\corchete{\fr{1}{2}\bI - \bD + \bS\fr{\bk}{D}}}} \bc = \bS\fr{\balpha}{D}  + \bc_{\infty},
}
and \eqref{eq:grad_discrete}
\eqn{
  \label{eq:grad_c2}
  \fr{1}{2} \bna \bc = \bna \bc_{\infty} + \bL \bw \bc + \bT \corchete{\fr{\bw \bk \bc}{D} - \fr{\balpha}{D}},
  %\fr{1}{2}\bna \bc =  \bna \bc_{\infty} + \bL \bc - \bT\corchete{-\fr{\bk\bc}{D} + \fr{\balpha}{D} },
}
where $\balpha$, $\bc_{\infty}$ and  $\bc$ are $N_b \times 1$ vectors and $\bk$ and $\bw$ are $N_b \times N_b$ diagonal matrices containing the value of the reaction rate $k$ and weight $w$ at each node. $\bna \bc $ and $\bna \bc_{\infty}$ are $3N_b \times 1$ vectors. 

Since only the tangential component of the surface gradient $\bna_{\parallel} \bc = \corchete{\bI-\bn\bn}\bna \bc = \bP_{\parallel}\bna \bc$ has an effect on the hydrochemical coupling through the slip velocity $\bu_s = \mu \bna_{\parallel} \bc$, we will only focus on the operator that relate  the tangential surface gradients with the forcing terms.\\
Substituting the solution $\bc = \bA^{-1}\corchete{\bS\fr{\bw\balpha}{D}  + \bc_{\infty}}$ into the tangential component of \eqref{eq:grad_c2} and rearranging gives
\eqn{
\bna_{\parallel} \bc = \underset{\bG_{\alpha}}{\underbrace{2\bP_{\parallel}\corchete{\pare{\bL\bw+\bT\fr{\bw\bk}{D}}\bA^{-1}\bS\bw-\bT}}}\fr{\balpha}{D} + \underset{\bG_{\infty}}{\underbrace{2\bP_{\parallel}\pare{\bL\bw+\bT\fr{\bw\bk}{D}}\bA^{-1}}}\bc_{\infty} + \underset{\bG_{\bna_\infty}}{\underbrace{2\bP_{\parallel}}}\bna \bc_{\infty}
}
where $\bG_{\alpha}$ and $\bG_{\infty}$ are $3N_b\times N_b$ matrices, and $\bG_{\bna_\infty}$ is $3N_b\times 3N_b$.
The solution can be written more compactly as 
\eqn{
\bna_{\parallel} \bc = \bG \cdot \underset{\bbf_{\bna}}{\underbrace{\corchete{\begin{array}{c}
      \fr{\balpha}{D}\\
      \bc_{\infty} \\
      \bna \bc_{\infty}
\end{array}}}}
}
where $\bG = \corchete{\bG_{\alpha},\, \bG_{\infty}\, \bG_{\bna_\infty}}$ is a $3N_b \times 5N_b$ matrix that gives the tangential surface gradients induced by the forcing terms in $\bbf_{\bna}$, namely the surface fluxes and background field. Its size depends on the number of grid points.\\
Note that the operator $\bG$ depends nonlinearly on $\bk$, which  might explain the  behaviour of the propulsion speed of a sphere with $N_b = 12$ near Da $\approx 4$ in Figure \ref{fig:damkohler}. 
For the sake of simplicity we will consider $\bk=\bzero$ hereafter.\\

The goal of grid optimization is to match the action of the operator $\bG_{ref}$ obtained with a reference grid, independently of a specific forcing $\bbf_{\bna}$. 
However, direct comparison between $\bG$ and $\bG_{ref}$ is impossible due to the difference in the number of nodes ($N_{b,ref}\gg N_b)$.
One alternative is to compare moments of $\bna_{\parallel} \bc$, such as the mean:  
\eqn{
\overline{\bna_{\parallel} \bc} = \overline{\bG} \cdot \bbf_{\bna}
}
where $\overline{\bG}$ is a $3\times 5N_b$ matrix given by 
\eqn{
\overline{\bG} = \bM \bG,
}
with $\bM$ the $3\times 5N_b$ average operator given by
\eqn{
\bM = \fr{1}{N_b}\corchete{\bI,\, \cdots,\, \bI}.
}
Then, using singular value decomposition 
$$\overline{\bG} = \bW^{\overline{\bG}}\bSigma^{\overline{\bG}}(\bV^{\overline{\bG}})^T,$$
one can define the relative error between the three singular values of $\overline{\bG}$ and $\overline{\bG}_{ref}$, as in Section \ref{sec:SVD}:
\eqn{E_{\overline{\bG}} =  \|\bSigma^{\overline{\bG}}_{ref} - \beta \bSigma^{\overline{\bG}}\|/\|\bSigma^{\overline{\bG}}_{ref}\|. }
The error on the mean, $E_{\overline{\bG}}$, and on higher moments, could either be minimized separately to optimize the grid for the Laplace problem only, or it could be added to the cost function \eqref{eq:cost_func}  to optimize the grid for the whole hydrochemical problem.

\bibliographystyle{elsarticle-num}
\bibliography{Biblio.bib}

\end{document}